\documentclass[lettersize,journal]{IEEEtran}
  \IEEEoverridecommandlockouts
\usepackage{verbatim}
\usepackage{cite}
\usepackage{amsmath,amssymb,amsfonts}
\usepackage{graphicx}
\usepackage{textcomp}
\usepackage{xcolor}
\usepackage{array}
\usepackage{epsfig}
\newtheorem{Remark}{\it Remark}[section]

\newtheorem{Proposition}{\it Proposition}[section]
\newtheorem{Lemma}{\it Lemma}[section]

\makeatletter
\newcommand{\Rmnum}[1]{\expandafter\@slowromancap\romannumeral #1@}
\makeatother
\usepackage{bm}
\usepackage{multicol}
\usepackage{setspace}
\usepackage{subfigure}
\usepackage{epstopdf}
\usepackage{fancyhdr} 
\usepackage{indentfirst}
\usepackage{enumerate}
\usepackage{stfloats}
\usepackage{bm}
\usepackage{multirow}
\usepackage{threeparttable}
\usepackage{CJK}
\usepackage[justification=centering]{caption}
\usepackage{makecell}
\usepackage{caption}
\usepackage{cases}
\usepackage{amsmath}
\usepackage{algorithm}
\usepackage{algpseudocode}

\usepackage{graphics}
\usepackage{caption}
\usepackage{booktabs}

\definecolor{deepblue}{rgb}{0.6,0,0.4}
\definecolor{red}{rgb}{0.2,0,0.6}

\def\BibTeX{{\rm B\kern-.05em{\sc i\kern-.025em b}\kern-.08em
		T\kern-.1667em\lower.7ex\hbox{E}\kern-.125emX}}
\begin{document}

 \title{Adaptive Source-Channel Coding for Semantic Communications }
\author{\IEEEauthorblockN{Dongxu Li, Kai Yuan, Jianhao Huang, Chuan Huang, Xiaoqi Qin, Shuguang Cui, and Ping Zhang}
\thanks{
  D. Li and K. Yuan are with the Shenzhen Future Network of Intelligence Institute, the School of Science and Engineering, and the Guangdong Provincial Key Laboratory of Future Networks of Intelligence, The Chinese University of Hong Kong, Shenzhen 518172, China (e-mail:dongxuli@link.cuhk.edu.cn and   kaiyuan3@link.cuhk.edu.cn).
  
  J. Huang is with the Department of Electrical and Electronic Engineering, the University of Hong Kong, Hong Kong 999077, China (email: Jianhaoh@hku.hk).

C. Huang and S. Cui are with the School of Science and Engineering, the Shenzhen Future Network of Intelligence Institute, and the Guangdong Provincial Key Laboratory of Future Networks of Intelligence, The Chinese University of Hong Kong, Shenzhen 518172, China (e-mail: huangchuan@cuhk.edu.cn and shuguangcui@cuhk.edu.cn).
  
X. Qin and P. Zhang are with the School of Information and Communication Engineering and the State Key Laboratory of Networking and Switching Technology,  Beijing University of Posts and Telecommunications, Beijing 100876, China (email: xiaoqiqin@bupt.edu.cn and  pzhang@bupt.edu.cn).

}
}

 \maketitle
 \begin{abstract}
   
   Semantic communications (SemComs) have emerged as a promising paradigm for joint
   data and task-oriented transmissions, combining the demands for both the bit-accurate delivery and end-to-end (E2E) distortion minimization. However, current joint source-channel coding (JSCC) in SemComs is not compatible with the existing communication systems and cannot adapt to the variations of the sources or the channels, while separate source-channel coding (SSCC) is suboptimal in the finite blocklength regime. To address these issues, we propose an adaptive source-channel coding (ASCC) scheme for SemComs over parallel Gaussian channels, where the deep neural network (DNN)-based semantic source coding and conventional digital channel coding are separately deployed and adaptively designed. To enable efficient adaptation between the source and channel coding, we first approximate the E2E data and semantic distortions as functions of source coding rate and bit error ratio (BER) via logistic regression, where BER is further modeled as functions of signal-to-noise ratio (SNR) and channel coding rate. Then, we formulate the weighted sum E2E distortion minimization problem for joint source-channel coding rate and power allocation over parallel channels, which is solved by the successive convex approximation.
    Finally, simulation results demonstrate that the proposed ASCC scheme outperforms typical deep JSCC and SSCC schemes for both the single- and parallel-channel scenarios while maintaining full compatibility with practical digital systems.

	\end{abstract}
	\begin{IEEEkeywords} 
	Semantic communications, task-oriented communications,  adaptive source-channel coding, power allocation, rate adaptation.
	\end{IEEEkeywords}
   \vspace{-0.5cm} 	
	\section{Introduction}
	The emergence of various intelligent applications, e.g., autonomous driving\cite{9679803} and virtual/augmented reality \cite{gunduz2022beyond}, has posed explosively growing demands for efficient and task-aware information delivery in wireless networks, making semantic communications (SemComs) a promising paradigm for the future 6G networks \cite{strinati20216g,huang2025visual,lin2023pushing,10529950}.
	 Unlike conventional communication systems that focus on accurate bit transmissions, SemComs prioritize extracting and transmitting the intrinsic information of data source for minimizing the end-to-end (E2E) distortion that is determined by specific tasks, e.g., object detection \cite{Qin} and data recovery\cite{balle2018variational}. This paradigm shift significantly improves transmission efficiency, particularly in the case with limited communication resources, by focusing on the task-relevant features and eliminating unnecessary contents for the considered tasks \cite{zhang2022toward}. 
	\begin{figure}[t]
	\centering
	\includegraphics[width=3
	in]{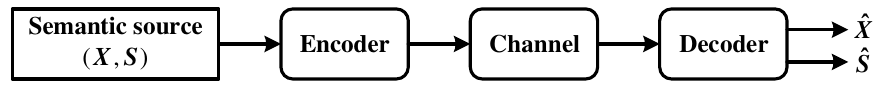}
	\caption{Framework of semantic communication systems.}
	\label{fig_system} 
    \end{figure}
    
   From an information-theoretic perspective, the framework of a typical SemCom is shown in Fig. \ref{fig_system}\cite{jcin_li,9844779,10465200, 10495330}. Specifically, the semantic source comprises two parts: an extrinsic observation $\bm{X}$ and an intrinsic semantic state $\bm{S}$, where $\bm{S}$ is typically not directly observable and can be inferred from $\bm{X}$\cite{9844779}. For instance, in image processing, raw image data represents extrinsic observation $\bm{X}$, while its extracted features for specific tasks, e.g., object classification \cite{jcin_li} and scene understanding \cite{gu2019survey}, correspond to intrinsic semantic state $\bm{S}$. In this framework,  SemComs primarily focus on encoding and transmitting observation $\bm{X}$ to simultaneously recover the reconstructed observation $\bm{\hat{X}}$ and semantic state $\bm{\hat{S}}$ at the receiver, ensuring both the data fidelity and semantic accuracy under the constraint of limited transmission resources. Specifically, there exist two main coding schemes for SemComs: separate source-channel coding (SSCC)\cite{jcin_li}, that performs semantic source compression and channel coding independently, and joint source-channel coding (JSCC) \cite{gunduz2024joint}, that directly maps semantic source data into channel input symbols. For discrete memoryless channels with infinite channel uses, theoretical analysis has shown that SSCC schemes can achieve optimality\cite{jcin_li,9844779}; however, for the cases with finite blocklength, JSCC schemes generally exhibit superior performance over SSCC ones since they can avoid performance loss caused by separate optimizations\cite{gunduz2024joint}.

With the development of deep learning techniques, deep neural networks (DNNs) \cite{8054694} have been actively explored in SemComs, offering potential solutions for JSCC implementation over practical data sources. The pioneering work in \cite{8723589} introduced a DNN-based JSCC scheme for image transmissions, which directly maps image pixels into complex-valued channel inputs. This approach was further extended to text transmissions using the Transformer architectures \cite{Qin2}, and its performance was enhanced through nonlinear transform coding for semantic feature extraction \cite{9791398}. However, these methods directly encode semantic source as analog symbols for transmissions, creating compatibility issues with the modern digital communication systems that are designed based on the idea of SSCC. To address this issue, recent studies in \cite{9252948,9998051,10495330} focused on incorporating quantization and constellation mapping modules into the DNN-based JSCC schemes. The authors in \cite{9998051} proposed DeepJSCC-Q, a JSCC scheme that employs a quantization layer to generate symbols with finite constellations for image transmissions. However, these approaches heavily rely on the E2E training for specific tasks and channel models\cite{10327757}; when the task or channel changes, the DNN-based JSCC schemes may need to be redesigned and retrained\cite{10495330,9252948}, leading to degraded generalization and adaptation capabilities.

Combining DNN-based semantic source coding with the digital channel coding techniques 
\cite{10327757,huang2024d,10521803}, e.g., low-density parity check (LDPC) and Polar codes \cite{8962344,gallager1968information,1057683}, offers a promising approach for maintaining compatibility with the existing digital systems while enabling close-to-optimal transmissions.  A rate-adaptive channel coding mechanism was developed in \cite{10327757} for multi-modal data transmissions, which allocates channel coding rates based on the semantic importance of different modalities.  However, this mechanism is specifically designed for multi-modal fusion tasks with a separate pre-trained semantic source encoder and does not achieve the source-channel adaptation.  
 Moreover,  the authors in \cite{huang2024d} explored joint optimization of DNN parameters for semantic source coding and the channel coding rate through E2E training, and the authors in \cite{10521803} addressed the resource allocation in Orthogonal Frequency Division Multiplexing (OFDM)-based SemComs by using reinforcement learning. However,  these approaches relied on online training or real-time model updates, which leads to extremely high computational complexity and latency\cite{gunduz2024joint}, making them less practical for dynamic communication scenarios.

	 This paper aims to address the fundamental challenge of establishing explicit relationships between semantic source coding and digital channel coding to enable efficient adaptive transmissions in SemComs. Specifically,
we consider a typical SemCom framework for parallel Gaussian channels, where the semantic source comprises an extrinsic observation data and the corresponding intrinsic semantic state\cite{jcin_li}. The observation data is first compressed by a DNN-based semantic source encoder and then protected by a group of digital channel encoders for transmissions over parallel channels to enable joint recovery of both the observation data and semantic state at the receiver.  Unlike the typical JSCC frameworks \cite{8054694,8723589,Qin2,9791398,9252948,9998051}, we adopt the SSCC structure where the source and channel coding are separately deployed, maintaining the compatibility with the existing communication systems. Unlike the conventional SSCC approaches, which typically adapt the channel coding only to the varying channel conditions\cite{cover1999elements}, our work reveals that for the considered SemCom system, both the E2E observation and semantic distortions are jointly determined by the source-channel coding rates and channel conditions, which implies the need for joint adaptation among the source and channel coding rates and power allocation to minimize E2E distortions. To enable precise adaptation, we derive the explicit E2E distortion models and propose an adaptive source-channel coding (ASCC) scheme to solve the joint optimization for the source-channel coding rates and power allocation over parallel channels.
	  The main contributions are summarized as follows:
	\begin{enumerate}
		\item $ \textbf{E2E Distortion Model}$:
		We characterize both the E2E observation and semantic distortions as functions of the source coding rate and BER caused by channel errors by using logistic regression. Then, BER is further approximated as functions of the received signal-to-noise ratio (SNR)  and channel coding rate: for ideal random coding, we exploit the finite blocklength transmission theory to derive the lower bound on BER; for practical channel coding schemes, e.g., Polar and LDPC codes, we apply the data regression method to obtain the BER approximations. By integrating the distortion and BER models,  we then derive explicit expressions for the E2E distortions by using key parameters: source and channel coding rates, power value, and channel coding scheme.
        \item $ \textbf{Model Selection-based ASCC Optimization} $: With the built distortion models, we formulate the weighted sum minimization problem for E2E observation and semantic distortions by jointly designing the source coding rate, channel coding rate, and power allocation.
 We propose a model selection-based approach to solve the above optimization problem: First, we construct a look-up table for $ N $ pre-trained DNN models, each of which corresponds to a specific source coding rate; next, for each DNN model, we develop a successive convex approximation (SCA)-based algorithm to solve the non-convex optimization for power allocation and channel coding rate adaptation, where we iteratively approximate and solve a sequence of convex versions of the original problem; finally, the optimal solution to the original problem is derived by searching the DNN model in the look-up table that achieves the minimum weighted sum distortions.
	\end{enumerate}
	

   The remainder of this paper is organized as follows. Section II describes the proposed ASCC scheme.  Section III builds the E2E distortion model. Section IV formulates the E2E distortion minimization problem.  Sections V and VI solve the formulated problems for the single- and parallel-channel cases, respectively.  Section VII shows the experimental results.  Finally, Section VIII concludes this paper.

  \emph{Notation}: Lowercase and uppercase letters, e.g., $ x $ and $ L $, denote scalars; Bold letters, e.g., $ \bm{x} $, denote vectors; $  \mathbb{E}(\cdot) $ denotes the expectation operator; $ \ln(x) $, $ \log_2(x) $, and $\log_{10}(x) $ denote natural, base-$2$, and base-$10$ logarithms, respectively; $e^{x} $ denotes the natural exponential function.

\section{System Model}

\begin{figure*}[htbp]  
	\centering
	\includegraphics[width=7in]{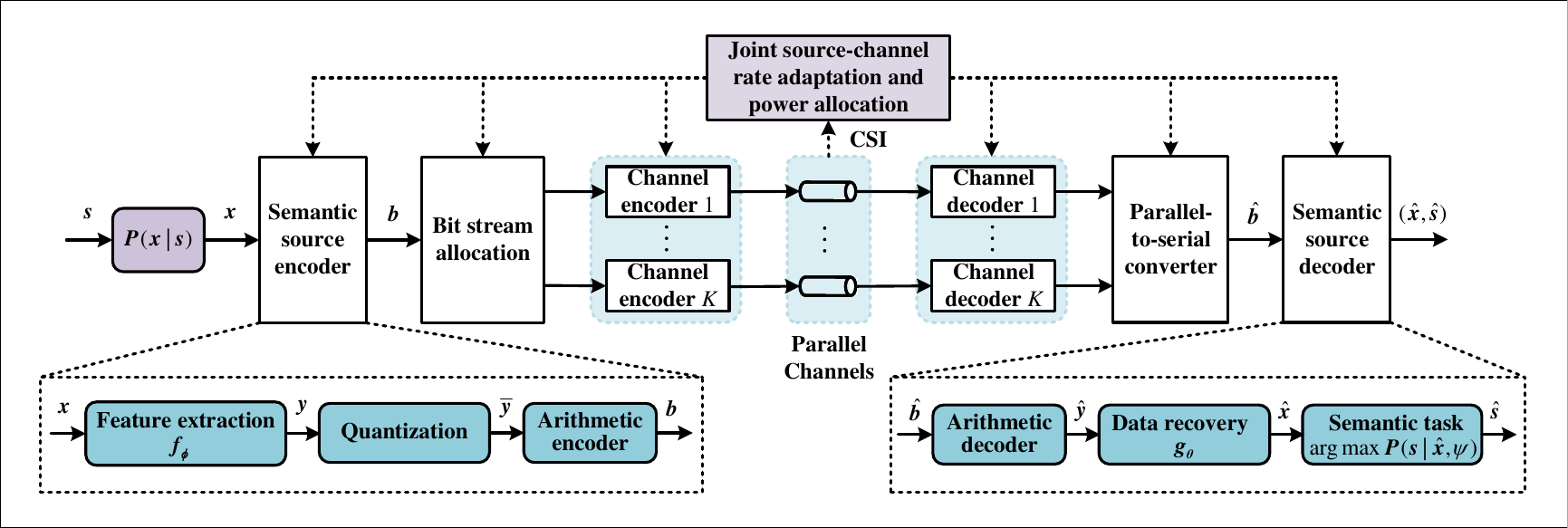}
	\caption{ASCC scheme over parallel Gaussian channels.}
	\label{fig_p_g_system} 
\end{figure*}

Consider a point-to-point digital SemCom system with $ K $ parallel Gaussian channels, denoted by a discrete set $ \mathcal{K} = \{ 1, \cdots, K \}$.  The semantic source is characterized by a joint probability distribution  $ (\bm{X}, \bm{S}) \sim P_{(\bm{X},\bm{S})} $, with $ \bm{X} \in \mathbb{R}^{M_1} $ being the extrinsic observation with dimension $M_1$ and $ \bm{S} \in \mathbb{R}^{M_2} $ being the intrinsic semantic state with dimension $M_2$\cite{jcin_li}. 
The SemCom system is about to transmit observation $ \bm{X} $ over $ K $ parallel Gaussian channels to the receiver for joint recovery of $ ( \hat{\bm{X}}, \hat{\bm{S}} ) $. To this end, we propose an ASCC scheme as shown in  Fig. \ref{fig_p_g_system}, consisting of a common semantic source encoder and $ K $ parallel channel encoders.

   \vspace{-0.3cm} 
\subsection{Transmitter}
Let $ (\bm{x},\bm{s}) $ denote the sample of the semantic source pair $ (\bm{X}, \bm{S})  $. As shown in Fig. \ref{fig_p_g_system}, the semantic source encoder directly processes observation sample $ \bm{x} $, where $  \bm{x}  $ is related to semantic state $ \bm{s} $ via conditional probability $ P(\bm{X}=\bm{x}|\bm{S} = \bm{s}) $.   First,  $\bm{x} $ is mapped to a $ W$-dimensional continuous-valued vector $ \bm{y} $ for lossy compression via a DNN-based feature extraction function $ f_{\phi} $ with parameter $ \phi $,   i.e., $  \bm{y} = f_\phi(\bm{x})$.  Then, $ \bm{y} $ is quantized as $\bar{\bm{y}} = \left\lfloor \bm{y} \right\rceil $, 
          where  $  \lfloor \cdot \rceil $ is the uniform scalar quantization operation (rounding to integers) \cite{9242247}.
           After that, a lossless entropy coding method, e.g., arithmetic coding \cite{10175391}, is applied to encode  $ \bar{\bm{y}} $ as bit stream $ \bm{b} $ based on its probability mass function $ P_{ \bar{\bm{Y}}}( \bar{\bm{Y}} = \bar{\bm{y}}) $, where $\bar{\bm{Y}}$ denotes the random variable corresponding to the quantized vector   $ \bar{\bm{y}} $.  Accordingly,  the expected source coding rate $R_{s}$ for encoding one source sample is defined as the entropy of $  \bar{\bm{Y}}$, i.e., $ R_{s} \triangleq  \mathbb{E} \{ -\log_2 P_{ \bar{\bm{Y}}} \} $.    

After source coding, each bit stream from one semantic source sample is allocated to a specific parallel channel, while bit streams from different source samples are distributed across the $K$ parallel channels for transmission. In other words, each bit stream can only be transmitted over one channel rather than being split across $K$ channels. 

Then, we consider a simple channel coding scheme where channel coding is performed independently over each of the parallel channels, rather than employing joint channel coding\footnote{Although joint coding across parallel channels has superior performance in the finite blocklength scenario, independent coding is more common in real-world parallel channel scenarios, e.g., multi-antenna systems\cite{tse2005fundamentals} and satellite communications\cite{elbert2008introduction} with multiple frequency bands, and can also achieve the same sum capacity as the blocklength approaches infinity\cite{tse2005fundamentals}.}. Specifically, consider the case that bit stream $ \bm{b} $, with a total length of $ B $, is allocated to the $ k $-th parallel channel, $k\in \mathcal{K} $. To facilitate the block channel coding, $ \bm{b} $ is first divided into $ T_k =  \lceil \frac{B}{N_k} \rceil $ blocks, where $ N_k $ 
  represents the number of data bits in each block and   $  \lceil \cdot \rceil $ denotes the round-up operation. Each block is then encoded as a symbol sequence of length $ L $, represented as $ [g_{k,t}^{(1)},\cdots,g_{k,t}^{(L)}] $ for block $t$, $t\in \{1,\cdots, T_k\} $, where $L$ denotes the number of symbols in each block.  Based on this independent encoding process, the channel coding rate of the $ k $-th parallel channel is calculated as $ R_{c,k} = \frac{N_k}{L} $.   Due to different channel state information (CSI) across the $K$ parallel channels, $R_{c,k}$ may vary with respect to $ k $, resulting in different average numbers of channel uses for transmitting source samples over each parallel channel. Specifically, the average number of channel uses per source sample transmitted over the $k$-th channel is $\frac{R_s}{R_{c,k}}$.  Additionally, the allocation ratio of source samples for the $ k $-th channel, which is proportional to its channel coding rate $ R_{c,k} $, is computed as $ \frac{R_{c,k}}{\sum_{k^\prime=1}^KR_{c,k^\prime}} $.  Hence, we calculate the average number of channel uses per source sample as 
            \begin{equation}
            	L_{\text{ave}} = \sum_{k=1}^K\frac{R_{c,k}}{\sum_{k^\prime=1}^KR_{c,k^\prime}}\cdot \frac{R_{s}}{R_{c,k}} = \frac{KR_{s}}{\sum_{k=1}^KR_{c,k}}. \label{L_ave}
            \end{equation}
 Moreover, the average bandwidth ratio (ABR) is defined as $ L_{\text{ave}}/M_1 $, measuring the average channel uses per dimension of observation $ X $.

   \vspace{-0.467cm}   	
\subsection{Channel Model}

  The input and output relationship at the $ t $-th block of the $ k $-th parallel channel is expressed as 
\begin{equation}
	\hat{g}_{k,t}^{(l)} =h_k\sqrt{P_k} g_{k,t}^{(l)}+ n_{k,t}^{(l)}, 
\end{equation}
 $l = 1,\cdots, L $, $ t = 1, \cdots,  T_k $, and $ k \in \mathcal{K}  $, where  $ g_{k,t}^{(l)} $ is the transmitted symbol with zero mean and unit average power for each  $ k \in \mathcal{K} $, $ \hat{g}_{k,t}^{(l)} $ denotes the corresponding received symbol,  $ n_{k,t}^{(l)} $ is independent and identically distributed (i.i.d.) circularly symmetric complex Gaussian (CSCG) noise with mean zero and variance $ \sigma^2 $, $ h_{k} \in \mathbb{C} $ is the constant complex channel coefficient of the $ k $-th parallel channel, and $ P_k $ is the allocated power for the $ k $-th parallel channel, satisfying the sum power constraint  
    \vspace{-0.2cm} 
\begin{equation}
    \sum_{k=1}^K P_k \leq P_{\text{max}}, \label{power_ineq}
\end{equation}
with $ P_{\text{max}} $ being the total power budget. Let $\bm{g}_{k}=[g_{k,1}^{(1)},\cdots,$
 $g_{k,1}^{(L)},\cdots,g_{k,T_k}^{(1)},\cdots,g_{k,T_k}^{(L)}] $ and $ \hat{\bm{g}}_{k}=[\hat{g}_{k,1}^{(1)},\cdots,\hat{g}_{k,1}^{(L)}, \cdots, \hat{g}_{k,T_k}^{(1)}$ $,\cdots, \hat{g}_{k,T_k}^{(L)}] $  denote the transmitted and received symbol sequences in the $k$-th parallel channel, respectively. The corresponding noise sequence is $ \bm{n}_{k}=[n_{k,1}^{(1)},\cdots,n_{k,1}^{(L)}, \cdots, n_{k,T_k}^{(1)}$ $,\cdots n_{k,T_k}^{(L)}] $. The received SNR in the $ k$-th parallel channel is $ \gamma_k = |h_k|^2P_k/\sigma^2 $.       

   \vspace{-0.4cm}   
\subsection{Receiver}
At the receiver, the received symbol sequence $ \hat{\bm{g}}_{k}$ of the $k$-th parallel channel is first decoded into bit stream $\hat{\bm{b}}$ by the $ k $-th channel decoder. Due to the channel noise, the recovered bit stream $ \hat{\bm{b}}$ might not be equal to the transmitted one.  We define $\rho_{b,k}$ as the bit error rate (BER) over the $k$-th parallel channel, which measures the average proportion of bits differing between the original bit stream $\bm{b}$ and recovered bit stream $\hat{\bm{b}}$. Then, all bit streams from the $ K $ channel decoders are merged into a single serial via a parallel-to-serial converter, waiting for the decoding at a common DNN-based semantic source decoder. The detailed semantic source decoding process is illustrated in Fig. \ref{fig_p_g_system}, where the recovered bit stream for semantic source sample $ (\bm{x}, \bm{s}) $ is denoted as $ \hat{\bm{b}} $. 
Specifically,  $ \hat{\bm{b}} $ is first decoded as the recovered feature vector $  \hat{\bm{y}} $ by the arithmetic decoder\cite{10175391}. Next, $ \hat{\bm{y}} $ is input into the DNN-based recovery function $ g_{\theta} $, parameterized by $ \theta $, to recover observation sample $ \bm{x} $. Finally,  semantic state $\bm{s}$ is recovered by applying the maximum a posteriori (MAP) scheme\cite{10175391}, i.e., \begin{equation}
	\hat{\bm{s}} = \arg \max\limits_{\bm{s}} P( \bm{s}|\hat{\bm{x}}, \psi),
\end{equation}where $ \hat{\bm{s}} $ is the recovered semantic state, and  $  P( \bm{s}|\hat{\bm{x}}, \psi) $ is the posterior probability estimated by a DNN, with parameter $ \psi $.

   \vspace{-0.4cm}  
\subsection{E2E distortion} \label{e2e_dis}
Accordingly, the recovered source sample $(\hat{\bm{x}},\hat{\bm{s}})$ is distorted by the lossy compression introduced by the DNN-based semantic source encoder and transmission errors caused by the channel noise. Here, we define the E2E observation and semantic distortions over the $ k $-th parallel channel as $ D_{o,k} \triangleq \mathbb{E}_{\bm{x},\bm{n}_k}{d_o(\bm{x},\hat{\bm{x}})} $ and $ D_{s,k} \triangleq \mathbb{E}_{\bm{s},\bm{n}_k}{d_s(\bm{s},\hat{\bm{s}})} $, respectively,
where  $ d_o(\cdot) $ and $ d_s(\cdot) $ denote the observation and semantic distortion measures, respectively. 
Moreover, following our preliminary work\cite{huang2024d},  $D_{o,k}$ and $ D_{s,k} $ can be modeled as the sum of source and channel distortions, respectively, i.e.,
\begin{equation}
	D_{w,k} \approx D_w^s(R_s) + D_w^c(R_s, \rho_{b,k}), \label{D_ok_app}
\end{equation}
where $ w \in \{ o,s \} $ is an index referring to either observation data or semantic state. Besides, $D_o^s(R_s)$ and $ D_s^s(R_s) $ represent the source distortions of the observation data and the semantic state, respectively, which vary with the source coding rate $R_s$ determined by the DNN parameters of the semantic source encoder; $D_o^c(R_s, \rho_{b,k})$ and $D_s^c(R_s, \rho_{b,k})$ denote the corresponding channel distortions of the observation data and the semantic state over the $k$-th parallel channel, respectively, which are jointly determined by $R_s$ and $\rho_{b,k}$. However, as discussed in \cite{huang2024d}, it is challenging to directly model channel distortions $ D_o^c(R_s, \rho_{b,k}) $ and $  D_s^c(R_s, \rho_{b,k})$ in closed-form expressions due to their complex dependence on both the unattainable Lipschitz constant of DNNs and BER $\rho_{b,k}$, where $\rho_{b,k}$ is jointly affected by channel coding scheme, channel coding rate, and SNR.

 \vspace{-0.3cm} 
 \section{E2E Distortion Modeling}

 In this section, we model the E2E observation distortion $ D_{o,k} $ and semantic distortion $ D_{s,k} $ in two steps: First, we approximately model both of them as functions of BER via logistic regression; second,
  we derive approximations for BER with respect to SNR and channel coding rate for various block channel coding schemes, including random coding \cite{5452208} and practical channel coding schemes\cite{8962344,gallager1968information,1057683}.
 
    \vspace{-0.5cm} 
\subsection{Distortion Modeling} \label{sec_dnn}
 Given the challenges in deriving closed-form expressions for E2E distortions discussed in Section \ref{e2e_dis}, we employ a data regression approach to model $D_{o,k}$ and $D_{s,k}$ effectively.
\subsubsection{Distortion measures}
Without loss of generality, we consider the widely studied image recovery and classification task as an example. Specifically, we treat the image itself as the observation state and its label information as the semantic state. For the observation distortion measure, we utilize the well-known mean squared-error distortion\cite{cover1999elements}, i.e.,
  \vspace{-0.15cm} 
\begin{equation}
	d_o(\bm{x},\hat{\bm{x}}) = \frac{1}{M_1} \|\bm{x}-\hat{\bm{x}} \|_2^2. \label{mse}
\end{equation}
    The semantic distortion measure is defined as the Hamming distortion\cite{cover1999elements}, i.e., 
    \vspace{-0.3cm}  
  \begin{equation}
  	d_s(\bm{s},\hat{\bm{s}}) = \begin{cases}
		   0, &  \mathrm{if} \ \bm{s}=\hat{\bm{s}},  \\
		1,  &  \mathrm{if} \  \bm{s} \neq \hat{\bm{s}}.
	\end{cases} \label{ds}
  \end{equation}
  Based on the above two distortion measures, we can compute the expected E2E distortions $D_{o,k}$ and $D_{s,k}$, respectively\footnote{It is worth noting that while we use specific distortion measures \eqref{mse} and \eqref{ds} in this work, other measures, e.g.,  structural similarity index (SSIM) \cite{huang2024d}  and task-specific metrics \cite{jcin_li}, can also be employed for various applications. Regardless of the specific choice, all these distortions can be effectively modeled by data regression in a similar way.}. 

\subsubsection{Approximate distortion model} 
  We adopt a data regression method to estimate distortions $ D_{o,k} $ and $ D_{s,k} $ for one image dataset, e.g., Caltech-UCSD Birds-$200$-$2011$ (CUB-$200$-$2011$)\cite{WahCUB_200_2011}.    To enable efficient semantic coding, we employ the hyperprior-based DNN model\cite{balle2018variational} for the feature extraction function $f_{\phi}$ and feature recovery function $g_{\theta}$ shown in Fig. \ref{fig_p_g_system}.
  With specifically designed training methods\cite{balle2018variational}, we obtain a set of $N$ DNN models corresponding to a discrete set of source coding rates $\mathcal{R}_s \triangleq \{R_{s,1},\cdots, R_{s,n},\cdots, R_{s,N}\}$\footnote{These DNN models are derived by training with different rate-distortion trade-offs, achieved by minimizing $\lambda D_o+R_s$ with various values of the hyperparameter $\lambda$, under error-free transmission conditions (i.e., $\bm{y} = \hat{\bm{y}}$).}, where $R_{s,n}$ is the source coding rate of the $n$-th DNN model.  Specifically, for each DNN model, we perform the following procedure: First, we use the DNN-based semantic source encoder to compress each image of the dataset to bit stream $ \bm{b} $. Next, to simulate the channel errors, recovered bit stream $ \hat{\bm{b}} $ at the receiver is obtained by randomly flipping some of the bits in $ \bm{b} $ according to a specific BER $ \rho_{b,k} $. Then, $ \hat{\bm{b}} $ is decoded by the semantic source decoder to recover both the image and its label. For each recovered sample, we compute the observation and semantic distortions by \eqref{mse} and \eqref{ds}.   Finally, $ D_{o,k} $ and $ D_{s,k} $ for each BER value and source coding rate $R_s$ are obtained by averaging the distortion values over all image samples.   
     \begin{figure}[t] 
\centering

\subfigure[ \label{fig_dis_o} $ \log_{10} D_{o,k} $]{
\includegraphics[width=1.68in]{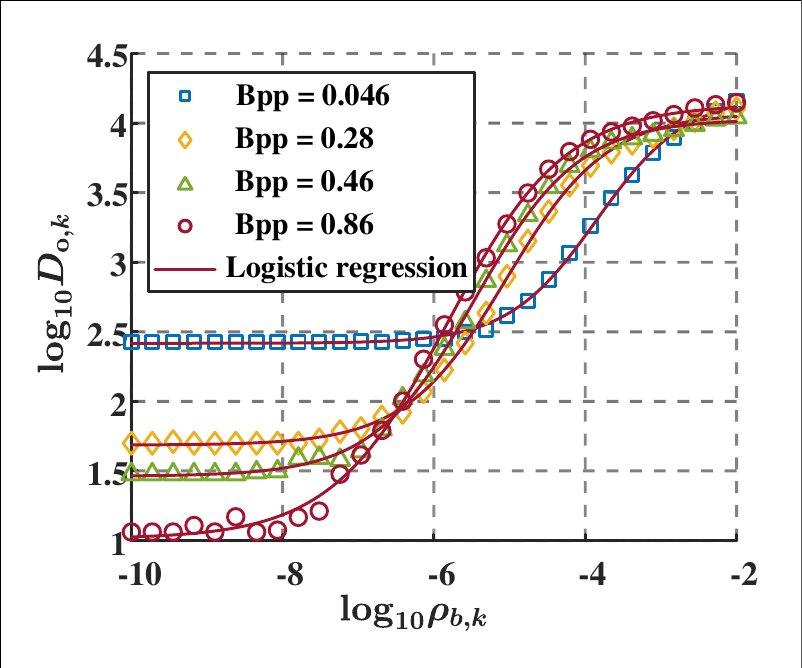}}
\hspace{-0.12in}
\subfigure[ \label{fig_dis_s}  $ D_{s,k} $ ]{
\includegraphics[width=1.69in]{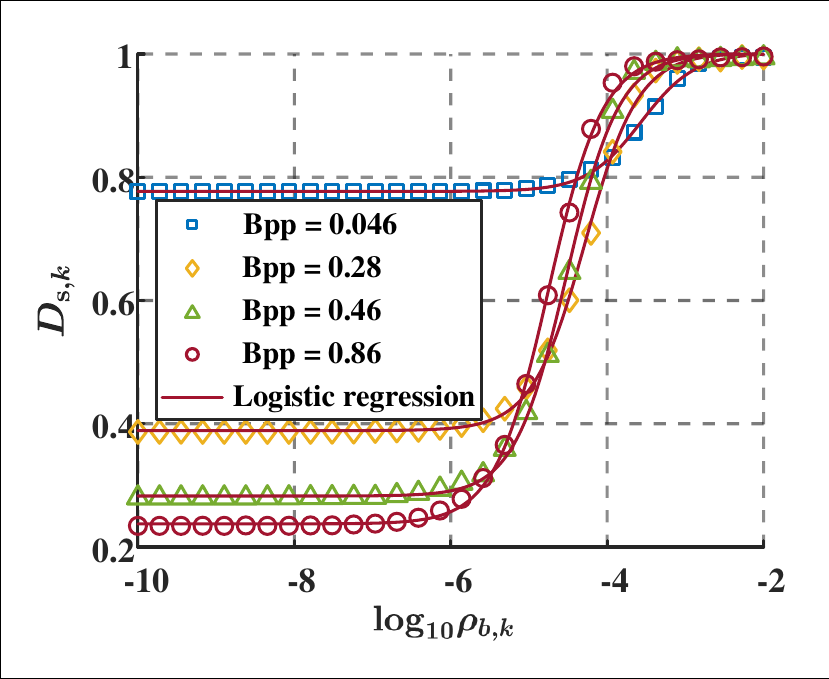}
}
\caption{E2E distortions $D_{o,k} $ and $ D_{s,k} $ as functions of BER for various values of $ R_s $. Here,  images in CUB-$200$-$2011$ are resized to $256 \times 256$ pixels, with the bit per pixel (Bpp) defined as $ \frac{R_{s}}{256 \times 256} $.}
\label{fig_varai_dis}
\end{figure}
   
 Based on the obtained observation and semantic distortion results, Fig. \ref{fig_varai_dis} plots the curves of $ \log_{10} D_{o,k} $ and $ D_{s,k} $ with respect to $ \log_{10} \rho_{b,k} $ and source rate $ R_s $, respectively,  with different $ R_s $'s corresponding to different DNN models.  Here, $D_{o,k} $ is expressed in the logarithmic scale to align with human perception of image distortion, while BER $ \rho_{b,k} $ is also represented in the logarithmic scale to capture its wide range.
 It is easy to see that both the $ \log_{10} D_{o,k} $ and $ D_{s,k} $ follow an ``S'' shape with respect to $ \log_{10} \rho_{b,k} $.  This observation motivates us to use the generalized logistic function to approximate $ \log_{10} D_{o,k} $ and $ D_{s,k} $ for any fixed $ R_s $, i.e.,
 \begin{align}
 	\log_{10} D_{o,k} &\approx \hat{D}_o(R_s,  \rho_{b,k}) \notag \\ &\triangleq
 	      \hat{D}_o^s(R_s)+\frac{\hat{D}_o^c(R_s)}{1+e^{-E_1^o(R_s) ( \log_{10} \rho_{b,k}-E_2^o(R_s))}}, \hspace{-0.25em}\label{Do_s}
 \end{align}
   \begin{align}
 	D_{s,k} &\approx D_s(R_s, \rho_{b,k}) \notag \\
 	     &\triangleq D_s^s(R_s)+\frac{D_s^c(R_s)}{1+e^{-E_1^s(R_s)(\log_{10}\rho_{b,k}-E_2^s(R_s))}}. \label{Ds_s}
 \end{align}
 Here,  $ \hat{D}_o^s(R_s) $, $ D_s^s(R_s) $, $ \hat{D}_o^c(R_s) $, $ D_s^c(R_s) $, $ E_1^o(R_s) $, $ E_1^s(R_s) $, $ E_2^o(R_s) $, and $ E_2^s(R_s) $ are all fitting parameters that need to be determined.    As shown in  Fig. \ref{fig_dis_o} and \ref{fig_dis_s}, the resulting generalized logistic functions \eqref{Do_s} and \eqref{Ds_s} accurately approximate $ \log_{10} D_{o,k} $ and $ D_{s,k} $, respectively.

 \begin{Remark}
As shown in Fig. \ref{fig_varai_dis}, based on the generalized logistic models \eqref{Do_s} and \eqref{Ds_s}, we observe that: 1) When BER is below certain thresholds, both the channel distortions in E2E distortions $D_{o,k} $ and $ D_{s,k} $  are close to zero, meaning that E2E distortions are primarily determined by their corresponding source distortions, respectively, with these thresholds varying across different tasks.  2) Increasing source coding rate $R_s$ reduces source distortions, while it simultaneously increases the corresponding channel distortions, i.e., higher $R_s$ makes the considered SemCom more sensitive to channel errors. 
\end{Remark}

These findings suggest a trade-off between the source coding rate $ R_s $ and BER $\rho_{b,k}$. Therefore, it is necessary to jointly optimize $R_s$ and $\rho_{b,k}$ to minimize the E2E distortions. 
   \vspace{-0.5cm}
 \subsection{BER Approximations}
 This subsection approximates BER $ \rho_{b,k} $ as a function of SNR $ \gamma_k $ and channel coding rate $ R_{k} $ over the $ k $-th parallel channel for both the random and practical channel coding schemes, respectively.  Random coding\cite{5452208} is an ideal channel coding scheme, which serves as the performance benchmark for the proposed ASCC framework, while practical channel coding schemes, e.g., Polar\cite{8962344} and LDPC codes \cite{gallager1968information,1057683}, are more attractive in practical communication systems.
 \subsubsection{Random coding} For the random coding scheme, we derive a lower bound on BER as follows.
 \begin{Proposition}
 	 For transmissions over the $k$-th Gaussian channel with SNR $\gamma_k$, blocklength $L$, and channel coding rate $R_{c,k}$, a lower bound on BER   for the random coding scheme is obtained as 
 	 \begin{equation}
\rho_{b,k} \geq \frac{Q\left(\sqrt{\frac{L}{V(\gamma_k)}} \left(C(\gamma_k)-R_{c,k}\right)\ln2 \right) }{R_{c,k}L},\ \forall k \in \mathcal{K}, \label{appro_rho}
\end{equation}
    where $ V(\gamma_k) $  is the channel dispersion satisfying $V(\gamma_k) = 1- \frac{1}{(1+\gamma_k)^2} $,  $C(\gamma_k) $ is the channel capacity defining as $ C(\gamma_k)= \log_2(1+\gamma_k) $, and the Q-function is defined as $ Q(x) = \int_x^\infty\frac{1}{\sqrt{2\pi}}e^{-\frac{t^2}{2}}dt $.
 \end{Proposition}
 \begin{IEEEproof} The block error probability for random coding scheme is approximated as \cite{5452208}
 	\begin{equation}
\epsilon_k = \epsilon(\gamma_k,R_{c,k}) =Q(\sqrt{L/V(\gamma_k)}(C(\gamma_k)-R_{c,k})\ln2). \label{eps}
\end{equation}Since the number of error bits in an erroneous block is at least 1, the minimum achievable BER is given by $\rho_{b,k}^{\min} = \frac{\epsilon_k}{R_{c,k}L}$. Therefore, we have derived the lower bound in \eqref{appro_rho}.
 \end{IEEEproof}

   \begin{figure}[t] 
\centering

\subfigure[ \label{fig_polar_snr_1} $ M = 4 $  ]{
\includegraphics[width=3 in]{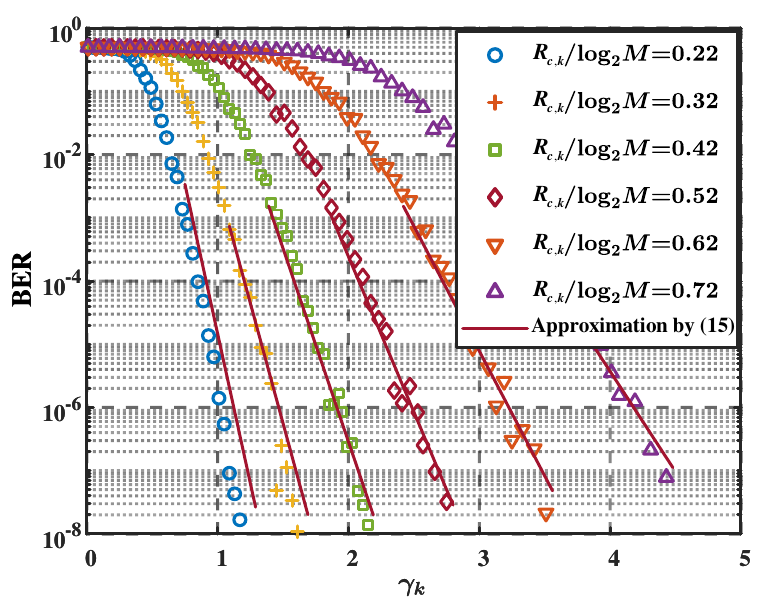}}
\subfigure[ \label{fig_polar_snr_2} $ M = 16 $]{
\includegraphics[width=3in]{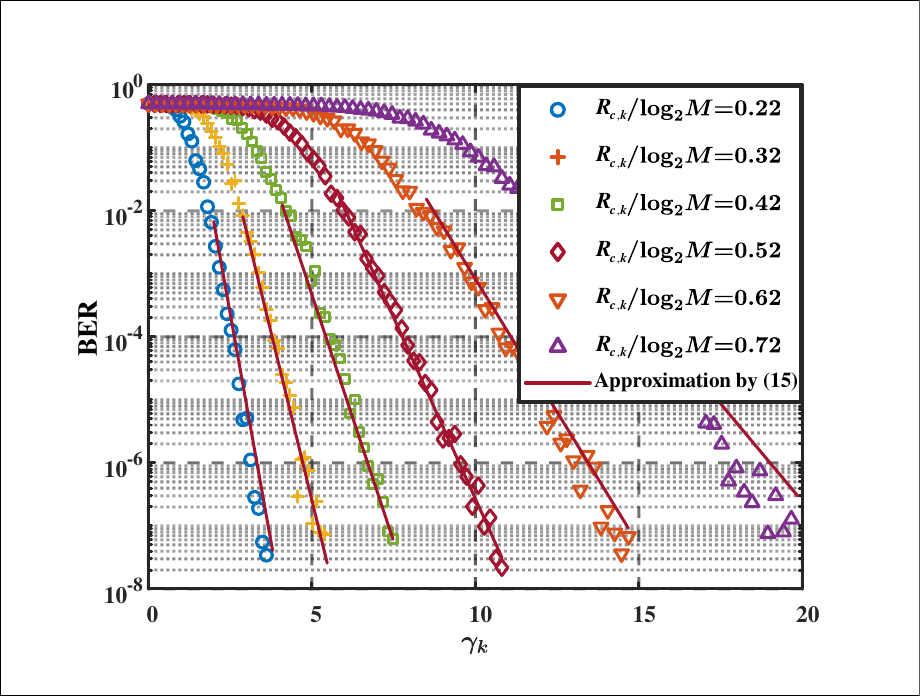}
}
\caption{BER vs. SNR $ \gamma_k $, channel coding rate $R_{c,k} $, and modulation order $ M $, with Polar code of length $ 256 $.}
\label{fig_polar_ber_final}
\end{figure}
  \subsubsection{Practical  coding}
For practical coding schemes, e.g., Polar and LDPC codes\cite{7998249}, there lack simple closed-form expressions for BER $ \rho_{b,k} $\footnote{There exist finite-length performance analysis methods for practical coding schemes, e.g., the finite-length scaling laws\cite{6866198} and density evolution approaches\cite{6823688}. However, these methods involve complex scaling exponents or asymptotic approximations that require complicated pre-computed parameters or iterative evaluations, and thus do not provide simple closed-form expressions for BER\cite{gallager1968information}.}.  To address this issue, similar to approaches in \cite{9306615}, we also apply the data regression method to model $ \rho_{b,k} $  as a function of SNR $ \gamma_k $ and channel coding rate $R_{c,k} $. Without loss of generality, we perform Monte Carlo simulations \cite{gu2019survey} for the  Polar code of length $256$. For each combination of the $M$-ary quadrature amplitude modulation ($M$-QAM) scheme and channel coding rate $R_{c,k}$, we simulate the transmissions of a large number of codewords over the $ k$-th parallel channel under different SNR values and compute the corresponding BER values.  The simulation results are shown in  Fig. \ref{fig_polar_ber_final}, which illustrates the variation of BER $ \rho_{b,k} $ with respect to SNR $ \gamma_k $ under various channel coding rates $R_{c,k} $ and  $M$-QAM schemes for Polar codes.

\begin{Remark} \label{Re_practical_code}
From Fig. \ref{fig_polar_ber_final}, we observe that: First, for a fixed channel coding rate $R_{c,k}$ and modulation order $M$, BER $ \rho_{b,k} $ in the logarithmic scale exhibits two distinct trends as SNR $\gamma_k$ increases: It decreases gradually when $ \rho_{b,k} > 10^{-3}$, and then decreases almost linearly when $ \rho_{b,k} \in [10^{-7}, 10^{-3}]$; for a fixed SNR and modulation order $M$, BER $ \rho_{b,k} $ increases monotonically as channel coding rate $R_{c,k}$ increases.
\end{Remark}

 As shown in Fig. \ref{fig_varai_dis}, when $\rho_{b,k} > 10^{-3}$, both the observation and semantic distortions reach their maximum values, indicating that the considered SemCom system cannot work over the region of $\rho_{b,k} > 10^{-3}$; when $\rho_{b,k} < 10^{-7}$, both the channel distortions of observation and semantic state in \eqref{D_ok_app} approach zero, which means that further BER reduction does not provide meaningful improvement for the considered SemCom system. Therefore, we only consider the case of $\rho_{b,k} \in [10^{-7}, 10^{-3}] $, and approximate $\log_{10} \rho_{b,k}$ as a linear function of SNR $\gamma_k$, i.e., \begin{equation}
	\log_{10} \rho_{b,k} = - \lambda\gamma_k + \mu, \label{fit_linear}
\end{equation}
where $ \lambda >0$ and $ \mu $ are the slope and intercept, respectively. Similar linear relationships can also be observed for other types of channel codes, e.g., LDPC\cite{gallager1968information}.

\begin{Remark}
Although practical channel codes typically exhibit waterfall-like BER curves with potential error floors at very low BER levels \cite{1057683}, e.g., $10^{-7}$ to $10^{-8}$, the proposed log-linear approximation for BER is specifically designed for and validated in the BER region of $[10^{-7}, 10^{-3}]$, which corresponds to the SemCom system's effective operating BER region shown in Fig. \ref{fig_varai_dis}. Moreover, as shown in Fig. \ref{fig_varai_dis}, it is easy to see that, beyond the region $[10^{-7}, 10^{-3}]$, the deviation between the real BER value and its approximation \eqref{fit_linear} almost does not affect the E2E observation and semantic distortions.
\end{Remark}

  \vspace{-0.05cm}
\begin{figure} 
\centering
\subfigure[ \label{fig_polar_a_1}  $ \ln \lambda $ vs.  $ R_{c,k}/\log_2{M} $   ]{
\includegraphics[width=1.7in]{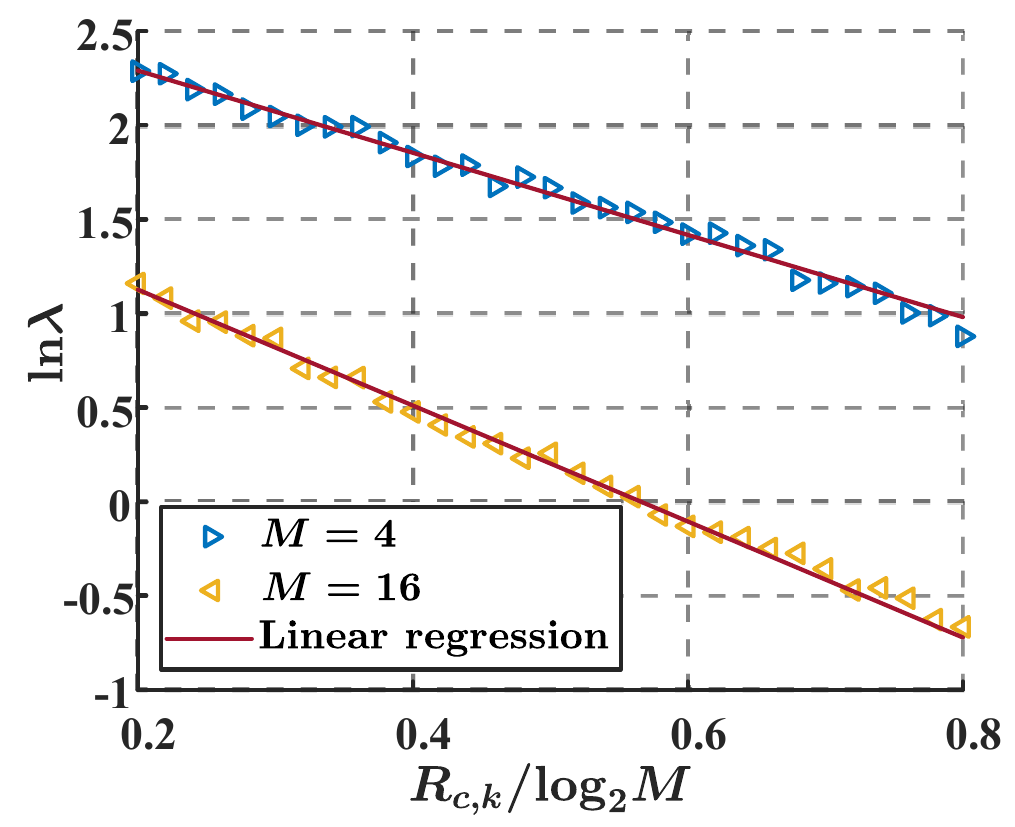}}
\hspace{-0.18in}
\subfigure[ \label{fig_polar_a_2} $ \mu $ vs.  $ R_{c,k} /\log_2{M}$]{
\includegraphics[width=1.7in]{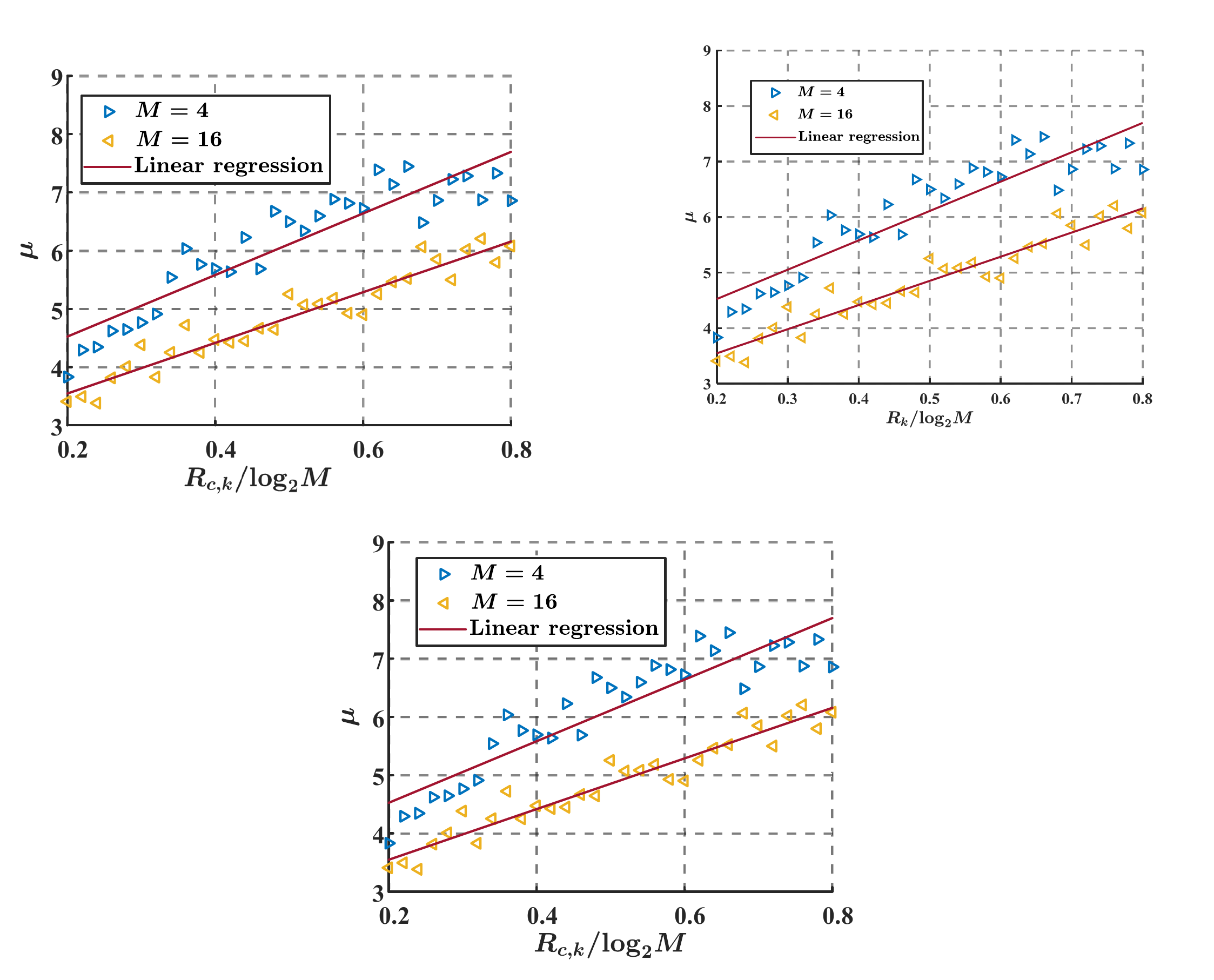}
}
  \vspace{-0.3cm}
\caption{Parameters  $ \lambda $ and $ \mu $ vs. channel coding rate $R_{c,k} $ and modulation order $ M $, with Polar code of length $256 $. }
\label{fig_polar_para_ber}
\end{figure}
It is noted that fitting parameters $ \lambda $ and $ \mu $ are jointly determined by the channel coding rate $ R_{c,k} $ and modulation order $ M $.  For each combination of  $ R_{c,k} $ and $ M $, the values of  $ \lambda $ and $ \mu $ are plotted in Fig. \ref{fig_polar_para_ber}. From this figure, it is easy to see that $ \lambda $ and $ \mu $ can be further modeled as 
\begin{equation}
 \ln \lambda = \lambda_1 R_{c,k}/\log_2{M}  + \lambda_2 , \label{fit_linear_1}
\end{equation}
\begin{equation}
	\mu = \mu_1 R_{c,k}/\log_2{M} + \mu_2, \label{fit_linear_2}
\end{equation}
where $ \lambda_1, \lambda_2, \mu_1 $, and $ \mu_2 $ are fitting parameters to be determined. Finally,
combining \eqref{fit_linear} with \eqref{fit_linear_1} and \eqref{fit_linear_2}, the modeled BER $ \rho_{b,k} $ for practical coding with $M$-QAM modulations is given as\begin{equation}
	\rho_{b,k} = 10^{-e^{\lambda_1 R_{c,k}/\log_2{M} + \lambda_2} \gamma_k + \mu_1 R_{c,k}/\log_2{M} + \mu_2}. \label{final_ber_appro}
\end{equation}
\begin{figure*}[bh]
\hrulefill
\begin{equation}
    D_{\text{ave}} \triangleq \sum_{k=1}^{K}\frac{R_{c,k}}{\sum_{k^{\prime}=1}^{K}R_{c,k^{\prime}}}\left(\alpha D_{o,k} + (1-\alpha) D_{s,k}\right) = \sum_{k=1}^{K}\frac{R_{c,k}\left(\alpha 10^{\hat{D}_o(R_s,\rho_{b,k})} + (1-\alpha) D_s(R_s,\rho_{b,k}) \right)}{\sum_{k^{\prime}=1}^{K}R_{c,k^{\prime}}}  \label{obj2}\end{equation}
\end{figure*}
To validate our proposed BER model, we compare the simulated BER values with those computed by \eqref{final_ber_appro} for different SNRs, channel coding rates, and coding schemes. The comparison results for Polar coding are shown in Fig. \ref{fig_polar_ber_final}, and the corresponding fitting parameters are summarized in Table \ref{table:1}. It is easy to see that the modeled BER in \eqref{final_ber_appro} accurately approximates the simulated BER values when BER is less than  $ 10^{-3}$.  Similar results can also be observed for the LDPC code.

%

\begin{table}[t]

\caption{Parameters $ \lambda_1 $, $ \lambda_2 $, $ \mu_1 $, and $ \mu_2 $ in \eqref{final_ber_appro} }
\footnotesize
\centering
\renewcommand{\arraystretch}{1.5} 
\setlength{\tabcolsep}{3pt}
\begin{tabular}{c|c|c|c|c|c|c} 
\Xhline{1.2pt} 
                  &  \begin{tabular}{@{}c@{}}Codeword\\[-0.5em]Length\end{tabular}               & Modulation  & $ \lambda_1 $ & $ \lambda_2 $ & $ \mu_1 $ &$ \mu_2 $  \\ \Xhline{1.2pt} 
\multirow{2}{*}{Polar} & \multirow{2}{*}{256} & QPSK & -2.1840 &   5.4558  &  5.2860  &  6.9342   \\ \cline{3-7} 
                  &                   & 16-QAM  & -3.0832    & 6.9736  &  4.3488 &  10.7088
 \\ \hline
\multirow{2}{*}{LDPC} & \multirow{2}{*}{4096} &   QPSK   & -2.4380 & 8.3714  & 14.1470 &  29.4932   \\ \cline{3-7} 
                  &                   &   16-QAM   & -3.1388  & 13.0368 &  12.2148  & 58.1672  \\ \Xhline{1.5pt} 
\end{tabular}
\label{table:1}
\end{table}
 \vspace{-0.4cm}

\section{Problem Formulation}
Based on the established distortion models \eqref{Do_s} and \eqref{Ds_s}, we now formulate our optimization problem for the proposed ASCC framework. In the considered SemCom system, the distortions $D_{o,k}$ and $D_{s,k}$ vary across the parallel channels due to different channel conditions. To evaluate the average performance, we define $D_{\text{ave}}$ as the expectation of the weighted sum of the E2E  distortions $D_{o,k}$ and $D_{s,k}$ over the parallel channels, i.e., \eqref{obj2}. Here, $ \frac{R_{c,k}}{\sum_{k^{\prime}=1}^{K}R_{c,k^{\prime}}} $ is the distortion weight for the $k$-th parallel channel, which is derived from \eqref{L_ave},  $\alpha \in [0,1]$ is the weight coefficient between observation distortion $D_{o,k}$ and semantic distortion $D_{s,k}$. By adjusting $\alpha$, SemCom system can tune the trade-off between these two distortions to meet different performance requirements.


Our goal is to minimize the average distortion $ D_{\text{ave}} $ in \eqref{obj2} by optimizing the source and channel coding rates and power allocation over $ K $ parallel channels, subject to constraints on the total transmitted power and the average channel uses.
By incorporating our derived distortion models and BER approximations, the optimization problem is formulated as
 \vspace{-0.1 cm}
\begin{align}
   	\text{(P1)}\min\limits_{\substack{R_s, \{P_k, R_{c,k} \}}}  &\eqref{obj2}  \notag\\
   \text{s.t.} \ \  \quad &\frac{KR_{s}}{\sum_{k=1}^KR_{c,k}} \leq L_{\text{max}}, \label{rate_ineq} \\ 
  & 0\leq R_{c,k} \leq C\left(\gamma_k\right),   P_k \geq 0,   \forall k \in \mathcal{K}, \label{big0}    \\
  & \eqref{power_ineq}, R_s \in \mathcal{R}_s, \label{all_constrain}  
\end{align}
where \eqref{rate_ineq} is the average channel uses constraint derived in \eqref{L_ave},  with $ L_{\text{max}} $ being the maximal number of channel uses. Here, we consider a practical case that the DNN model for semantic coding is not trained online to minimize \eqref{obj2} while selected from a look-up table with source coding rate $R_s$ in discrete set $ \mathcal{R}_s$. The look-up table\footnote{To reduce the storage requirements for this look-up table, a variable-rate source coding method can be employed \cite{kamisli2024variable}, which enables a single unified DNN model to support multiple rate-distortion operating points through a configurable scalar input parameter.} stores $ N $ pre-trained DNN models derived in Section \ref{sec_dnn},  each characterized by its source coding rate $R_s \in \mathcal{R}_s$, DNN parameters $\{\phi,\theta,\psi\}$, and the corresponding fitting parameters in $\hat{D}_o(\cdot)$ and $D_s(\cdot)$.

To efficiently solve Problem (P1), we propose a model selection-based optimization approach that decouples the original problem into the selection of source coding rate $R_s$ (and its corresponding DNN model configuration) and optimization of power allocation variables $\{P_k\}$ and channel coding rates $\{R_{c,k}\}$. The main steps are given as follows:
\begin{enumerate}
	\item $ \textbf{Resource allocation with fixed DNN model}$: For each pre-trained DNN model in the look-up table, Problem (P1) is simplified into a power allocation and channel coding rate adaptation problem, which is given as 
 \vspace{-0.1 cm}
	  \begin{eqnarray}
   	\text{(P2)} &  \min\limits_{\{P_k, R_{c,k} \} } & \eqref{obj2}  \notag \\
  & \text{s.t.} & \eqref{power_ineq},  \eqref{rate_ineq},  \eqref{big0}. \label{constrain_p2}  \end{eqnarray}
  We solve Problem (P2) for each pre-trained model to obtain their respective minimum objective values.
 \item $ \textbf{Find the optimal DNN model}$:  The solution to Problem (P1) is then derived by selecting the model in the look-up table that achieves the lowest objective value in \eqref{obj2}.
\end{enumerate}

 In our proposed approach, all DNN models can be pre-trained offline, which eliminates the need for online model training and significantly reduces the computational burden. However, it is noted that Problem (P2) remains non-convex and difficult to solve due to the complicated expression in \eqref{obj2}. In the following sections, we propose efficient algorithms to solve Problem (P2) for different channel coding schemes.



 \vspace{-0.27 cm}
\section{Single Gaussian Channel}
In this section, we consider the single channel scenario, i.e., $ K=1 $, to analyze the optimal power and channel coding rate design in Problem (P2). For convenience of analysis, we drop the channel index $k$ for all variables. Then, Problem (P2) is rewritten as 	   \begin{eqnarray}
   	\text{(P3)} & \min\limits_{P, R_c }   & \alpha 10^{\hat{D}_o(R_s,\rho_b)} + (1-\alpha) D_s(R_s,\rho_b) \label{obj_1}  \\
   	 & \text{s.t.} & 0\leq P \leq P_{\text{max}},  \label{P_1} \\
    &   & \frac{R_s}{L_{\text{max}}} \leq R_c < C(\gamma), \label{R_1} 
   \end{eqnarray}
   where \eqref{obj_1} is obtained from \eqref{obj2}  by setting $ K=1 $, and \eqref{R_1} is derived from \eqref{rate_ineq} and \eqref{big0}. It is easy to see that  Problem  (P3) achieves its optimum value at $ P = P_\text{max} $. Moreover, from \eqref{Do_s} and \eqref{Ds_s}, $ \hat{D}_o(\cdot) $ and $ D_s(\cdot) $  are both monotonically increasing as $ \rho_b$ increases.  Therefore, Problem (P3)  can be simplified as
  	   \begin{eqnarray}
   	\text{(P3.1)} & \min\limits_{ R_c}   & \rho_b =  \eqref{appro_rho} \  \text{or} \  \eqref{final_ber_appro} \label{obj_rho_b} \\
   		 & \text{s.t.} & R_{s} /L_{\text{max}} \leq R_c \leq C\left(\gamma_{\text{max}}\right), \label{Pb_3} 
   \end{eqnarray}
   where $ \gamma_{\text{max}} $ is defined as $\gamma_{\text{max}} \triangleq |h|^2P_\text{max}/\sigma^2 $.
   The following proposition gives the optimal solution to Problem (P3.1).
   \begin{Proposition} \label{Prpo_R}
   	Given $   \frac{R_{s}}{L_{\text{max}}}\leq C\left(\gamma_{\text{max}}\right)$, for the random coding case,  the optimal solution of Problem (P3.1) is $ R_c^{\star}= \frac{R_{s}}{L_{\text{max}}} $ if $ \frac{R_{s}}{L_{\text{max}}} \geq \frac{\sqrt{2 \pi}\log_2 e}{2 \sqrt{L}}  $; for the practical coding case, $ R_c^{\star}= \frac{R_{s}}{L_{\text{max}}} $ is the optimal solution of Problem (P3.1).
   \end{Proposition}
   \begin{IEEEproof}
   	Please see Appendix \ref{ap_prpo_R}.
   \end{IEEEproof}
   
  \begin{Remark}
  	 From Proposition  \ref{Prpo_R}, the optimal solution of Problem (P3.1) requires $ \frac{R_{s}}{L_{\text{max}}} \geq \frac{\sqrt{2 \pi}\log_2 e}{2 \sqrt{L}} $ for random coding. This condition is typically satisfied in practical communication systems since the threshold $\frac{\sqrt{2 \pi}\log_2 e}{2 \sqrt{L}}$ is negligibly small for practical blocklengths (e.g., $ \frac{\sqrt{2 \pi}\log_2 e}{2 \sqrt{L}} < 0.12$ for $L \geq 256$). Therefore, we consider the case that   $ R_c^{\star}= \frac{R_{s}}{L_{\text{max}}} $ always holds in the following analysis. 
  \end{Remark}
  
\section{Parallel Gaussian Channels}
In this section, we solve Problem (P2) for the general parallel channel case, i.e., $K \geq 2$. First, we simplify Problem (P2) by leveraging the optimal channel coding rate derived in Proposition  \ref{Prpo_R}. Then, we propose an SCA-based algorithm to efficiently solve Problem (P2).

\subsection{Problem Simplification}

Based on the results obtained by the single-channel case, the following proposition regarding the sum of optimal channel coding rates is derived for the parallel Gaussian channels.
 \begin{Proposition} \label{Prop_total_rate}
 	The optimal channel coding rates $ R_{c,1}^\star, \cdots, R_{c,K}^\star $ to Problem (P2) satisfy $ \sum_{k=1}^K R_{c,k}^\star =  \frac{KR_{s}}{L_{\text{max}}}$. 
 \end{Proposition}
 \begin{IEEEproof}
 Please see Appendix \ref{ap_Prop_total_rate}.
 \end{IEEEproof}
 
Using Proposition \ref{Prop_total_rate}, Problem (P2) is simplified as
    	   \begin{align}
    	      	\text{(P4)}   \min\limits_{ \{P_k, R_{c,k} \} } & \sum_{k=1}^{K}\frac{L_{\text{max}} R_{c,k}}{KR_{s}}\Big(\alpha 10^{\hat{D}_o(R_s,\rho_{b,k})}  \notag \\ &+(1-\alpha) D_s(R_s,\rho_{b,k}) \Big)  \label{obj_ori_simp}  \\
    	      	   \text{s.t.} \ \ \  & \eqref{constrain_p2}, \notag\end{align}
  where \eqref{obj_ori_simp} is obtained by replacing  $ \sum_{k=1}^K R_{c,k}$ in \eqref{obj2} with $ \frac{KR_{s}}{L_{\text{max}}}$.  

 \vspace{-0.5 cm}
   \subsection{Random Coding Case} 

   In this subsection, we utilize the SCA method to approximately solve Problem (P4) for the random coding case.  The main steps of the SCA method are given as follows:
   \subsubsection{Introducing auxiliary variables}
  By introducing auxiliary variables  $ r_k = \log_{10} R_{c,k} $, $ \hat{\rho}_{b,k} = \log_{10}\rho_{b,k} $, $ d_{o,k} $ and $ d_{s,k} $,  $ \forall  k \in\mathcal{K}$, and denoting $ \bm{\nu}_k = [r_k, \hat{\rho}_{b,k}, d_{o,k}, d_{s,k},  P_k, R_{c,k}] $ as the new design variable vector, 
 Problem (P4) for the random coding case is transformed as
 \begin{figure*}[bh]
\hrulefill
  \vspace{-0.18cm}
\begin{equation}
\frac{L_{\text{max}}}{KR_{s}}\sum_{k=1}^{K} \alpha 10^{r_k+e^{d_{o,k}}+\hat{D}_o^s(R_s)} + (1- \alpha ) (D_s^s(R_s)R_{c,k} + e^{d_{s,k}})   \label{obj_simple_0}
    \end{equation}
     \setcounter{equation}{37}
                \vspace{-0.2cm}
    \begin{equation}
    	  	U^{(i)}_{1,k}( \gamma_k, R_{c,k}) =  - \hat{a}(\hat{w}_{k}^{(i)})\left(\sqrt{L} \left(C(\gamma_k)-R_{c,k}\right)\ln2-\hat{w}_{k}^{(i)} \right)\log_{10}e- \log_{10}(R_{c,k}L)  + \log_{10} Q(\hat{w}_{k}^{(i)}) \label{convex_u1}
    \end{equation}
         \setcounter{equation}{42}
           \vspace{-0.2cm}
    \begin{equation}
    	   	U^{(i)}_{3,k}(\hat{\rho}_{b,k}, d_{o,k}) = - e^{d_{o,k}^{(i)}-E_1^o(R_s) (\hat{\rho}_{b,k}^{(i)}-E_2^o(R_s))}(d_{o,k}-d_{o,k}^{(i)}-E_1^o(R_s)(\hat{\rho}_{b,k}-\hat{\rho}_{b,k}^{(i)})+1)- e^{d_{o,k}^{(i)}}(d_{o,k}-d_{o,k}^{(i)}+1)+ \hat{D}_o^c(R_s)\label{convex_u3}
    \end{equation}
             \vspace{-0.2cm}
    \begin{equation}
    	  	U^{(i)}_{4,k}(\hat{\rho}_{b,k}, d_{s,k}, R_{c,k}) = -e^{d_{s,k}^{(i)}-E_1^s(R_s) (\hat{\rho}_{b,k}^{(i)}-E_2^s(R_s))}(d_{s,k}-d_{s,k}^{(i)}-E_1^s(R_s)(\hat{\rho}_{b,k}-\hat{\rho}_{b,k}^{(i)})+1) - e^{d_{s,k}^{(i)}}(d_{s,k}-d_{s,k}^{(i)}+1) + D_s^c(R_s)R_{c,k}  \label{convex_u4}
    \end{equation}
\end{figure*}
     \setcounter{equation}{27}
\begin{align}
   	\text{(P5.1)} \ \min\limits_{ \left\{ \bm{\nu}_k\right\} } \ & \eqref{obj_simple_0} \notag \\
   \text{s.t.} \;\;  &  \log_{10} \Bigg(\frac{Q\left(\sqrt{L} \left(C(\gamma_k)-R_{c,k}\right)\ln2 \right) }{R_{c,k}L}\Bigg)  \leq \hat{\rho}_{b,k},   \label{rho_b_ineq}   \\
   \quad &  \frac{\hat{D}_o^c(R_s)}{1+e^{-E_1^o(R_s) (\hat{\rho}_{b,k}-E_2^o(R_s))}} \leq e^{d_{o,k}},  \label{D_oi_ineq}  \\
  \quad &   \frac{D_s^c(R_s)R_{c,k}}{1+e^{-E_1^s(R_s) (\hat{\rho}_{b,k}-E_2^s(R_s))}}\leq e^{d_{s,k}}, \label{D_si_ineq} \\ 
   \quad & \log_{10}R_{c,k}\leq r_k ,  \label{log_r}  \\
   \quad &  0< R_{c,k} \leq C(\gamma_k), \    P_k > 0, \  \forall k \in \mathcal{K}, \label{R_bound} \\
    \quad & \eqref{power_ineq},\eqref{rate_ineq},    \label{R_constraints}  \end{align} 
 where \eqref{obj_simple_0} and \eqref{D_oi_ineq}-\eqref{log_r} are obtained by introducing auxiliary variables $ \hat{\rho}_{b,k} $,  $ d_{o,k} $, and $ d_{s,k} $ into \eqref{Do_s}, \eqref{Ds_s}, and \eqref{obj_ori_simp}. Besides, \eqref{rho_b_ineq} is derived from \eqref{appro_rho} by approximating $ V(\gamma_k) $ as $1$\cite{8933345,8253477}.
  Additionally, \eqref{R_bound} is obtained from \eqref{big0} by replacing $ R_{c,k} \geq 0 $  with $ R_{c,k} > 0 $ to ensure that auxiliary variable $ r_k =\log_{10} R_{c,k} $ is well defined. Note that for the optimal solution to Problem (P5.1), all constraints in \eqref{rho_b_ineq}-\eqref{log_r} must hold with equality; otherwise, the objective value in \eqref{obj_simple_0} can be further reduced by decreasing $ \hat{\rho}_{b,k} $, $ e^{d_{o,k}} $, $ e^{d_{s,k}} $, or $ r_k $. It is easy to see that the transformation from Problem (P4) to Problem (P5.1) preserves equivalence for all feasible solutions except the case $R_{c,k} = 0$. Furthermore, the solution to Problem (P5.1) can approach arbitrarily close to $R_{c,k} = 0$ as $r_k \to -\infty$.

 Additionally,  
  inequalities \eqref{D_oi_ineq} and \eqref{D_si_ineq} can be equivalently rewritten as
  \begin{equation}
     - e^{d_{o,k}-E_1^o(R_s) (\hat{\rho}_{b,k}-E_2^o(R_s))}- e^{d_{o,k}}+ \hat{D}_o^c(R_s) \leq 0,\label{new_d_oi_ineq}
  \end{equation}
   \begin{equation}
    - e^{d_{s,k}-E_1^s(R_s) (\hat{\rho}_{b,k}-E_2^s(R_s))} - e^{d_{s,k}} +	 D_s^c(R_s)R_{c,k} \leq 0. \label{new_d_si_ineq}
   \end{equation}
Problem (P5.1) is then equivalent to the following problem 
\begin{eqnarray}
   	\text{(P5.2)} & \min\limits_{ \left\{ \bm{\nu}_k\right\} }  & \eqref{obj_simple_0}   \notag  \\
  & \text{s.t.}  & \eqref{rho_b_ineq}, \eqref{log_r}-\eqref{new_d_si_ineq}. \end{eqnarray}
  Note that the left-hand sides (LHSs) of inequalities \eqref{rho_b_ineq}, \eqref{log_r}, \eqref{new_d_oi_ineq}, and \eqref{new_d_si_ineq} are all non-convex functions, thus making  Problem (P5.2) non-convex.     
  \subsubsection{Constraints approximation} We utilize the SCA method to deal with the non-convex constraints in \eqref{rho_b_ineq}, \eqref{log_r}, \eqref{new_d_oi_ineq}, and \eqref{new_d_si_ineq} to iteratively solve Problem (P5.2), thereby approximating it as a series of convex subproblems. Specifically, in the $ i $-th iteration  ($i>0 $), consider the current value of the design variable vector in Problem (P5.2) is denoted as $\bm{\nu}_k^{(i)} = [r_k^{(i)},d_{o,k}^{(i)},d_{s,k}^{(i)},\hat{\rho}_{b,k}^{(i)},P_k^{(i)},R_{c,k}^{(i)}] $, $ \forall k \in\mathcal{K} $. We first derive a convex upper bound for the LHS of \eqref{rho_b_ineq} in the following proposition.
  \begin{Proposition} \label{Prop_up}
  	The LHS of \eqref{rho_b_ineq} is upper bounded as
  \begin{equation}
  	 \log_{10} \left(\frac{Q\left(\sqrt{L} \left(C(\gamma_k)-R_{c,k}\right)\ln2 \right) }{R_{c,k}L}\right)  \leq U^{(i)}_{1,k}( \gamma_k, R_{c,k}), \label{upper1}
  \end{equation}
  where $ U^{(i)}_{1,k}( \gamma_k, R_{c,k}) $ is expressed as \eqref{convex_u1}, $\hat{w}_{k}^{(i)} $ is defined as  $ \hat{w}_{k}^{(i)} = \sqrt{L} (C(|h_k|^2P_k^{(i)}/\sigma^2)-R_{c,k}^{(i)})\ln2  $, and $ \hat{a}(\hat{w}_{k}^{(i)}) $ satisfies $  \hat{a}(\hat{w}_{k}^{(i)}) = e^{-\frac{(\hat{w}_{k}^{(i)})^2}{2}}/(\sqrt{2\pi}Q(\hat{w}_{k}^{(i)})) $.  
  \end{Proposition}
  \begin{IEEEproof}
  	Please see Appendix \ref{ap_Prop_up}.
  \end{IEEEproof}

 Although LHSs of inequalities  \eqref{log_r}, \eqref{new_d_oi_ineq}, and \eqref{new_d_si_ineq} are all non-convex, they are all sums of convex and concave functions.
   By replacing the concave functions in \eqref{log_r},  \eqref{new_d_oi_ineq},  and \eqref{new_d_si_ineq}  with their first-order  Taylor expansions, global convex upper bounds can be derived for the LHSs of  \eqref{log_r}, \eqref{new_d_oi_ineq}, and \eqref{new_d_si_ineq}  as  
    \setcounter{equation}{38}
      \begin{equation}
   	 \log_{10}R_{c,k} \leq U^{(i)}_{2,k}( R_{c,k}),
   \end{equation}
   \begin{align}
    &- e^{d_{o,k}-E_1^o(R_s) (\hat{\rho}_{b,k}-E_2^o(R_s))}  - e^{d_{o,k}} + \hat{D}_o^c(R_s) \notag \\ &\leq U^{(i)}_{3,k}(\hat{\rho}_{b,k}, d_{o,k}),
   \end{align}
   \begin{align} &- e^{d_{s,k}-E_1^s(R_s) (\hat{\rho}_{b,k}-E_2^s(R_s))} - e^{d_{s,k}} + D_s^c(R_s)R_{c,k} \notag \\ & \leq U^{(i)}_{4,k}(\hat{\rho}_{b,k}, d_{s,k}, R_{c,k}), 
   \end{align}	
 where $ U^{(i)}_{2,k}(\hat{\rho}_{b,k}, d_{o,k}) $,  $ U^{(i)}_{3,k}(\hat{\rho}_{b,k}, d_{s,k}, R_{c,k}) $,  and $ U^{(i)}_{4,k}( r_k, R_{c,k}) $ are given as
      \begin{equation}
   	U^{(i)}_{2,k}( R_{c,k}) = \frac{R_{c,k}-R_{c,k}^{(i)}}{R_{c,k}^{(i)}\ln10}  + \log_{10}R_{c,k}^{(i)}, \label{convex_u2}
   \end{equation}
   \eqref{convex_u3}, and \eqref{convex_u4}, respectively.

   By replacing the LHSs of \eqref{rho_b_ineq}, \eqref{log_r}, \eqref{new_d_oi_ineq}, and \eqref{new_d_si_ineq} with their convex upper bounds \eqref{convex_u1}, \eqref{convex_u2}, \eqref{convex_u3}, and \eqref{convex_u4}, respectively, we have the following convex subproblem in the $i$-th iteration, i.e.,
    \setcounter{equation}{45}
       \vspace{-0.2cm}
\begin{eqnarray}
   	\text{(P5.2.}i\text{)}  &\min\limits_{ \left\{ \bm{\nu}_k\right\} }  & \eqref{obj_simple_0}   \notag  \\
  & \text{s.t.} & U^{(i)}_{1,k}( \gamma_k, R_{c,k}) \leq \hat{\rho}_{b,k}, \ \forall k \in \mathcal{K},  \\
    & & U^{(i)}_{2,k}( R_{c,k}) \leq r_k, \ \forall k \in \mathcal{K}, \label{u2_ineq} \\
  & & U^{(i)}_{3,k}(\hat{\rho}_{b,k}, d_{o,k}) \leq 0, \ \forall k \in \mathcal{K},  \label{u3_ineq}\\ 
  & & U^{(i)}_{4,k}(\hat{\rho}_{b,k}, d_{s,k}, R_{c,k}) \leq 0, \ \forall k \in \mathcal{K},  \label{u4_ineq} \\ 
  &   & \eqref{R_bound}, \eqref{R_constraints},
  \end{eqnarray}
  which can be efficiently solved by some optimization tools, e.g., CVX\cite{boyd2004}.  Moreover, it is easy to see that the optimal value of Problem (P5.2.$i$) serves as an upper bound for Problem (P5.2). The SCA-based algorithm, which iteratively solves Problem (P5.2.$i$) to tackle Problem (P5.2), is summarized in Algorithm  \ref{abel_1}. The convergence analysis of this algorithm is discussed in \cite{Beck}, showing that it converges to a suboptimal solution to Problem (P5.2) that satisfies the KKT conditions.

     \setcounter{equation}{54}
 \begin{figure*}[bh]
 \hrulefill
   \vspace{-0.18cm}
  \begin{equation}
 	U^{(i)}_{5,k}(p_k, R_{c,k}) = -\frac{1}{\sigma^2}|h_k|^2e^{p_k^{(i)} + \frac{\lambda_1 R_{c,k}^{(i)}}{\log_2 M} + \lambda_2}\left(p_k-p_k^{(i)} + \frac{\lambda_1(R_{c,k}-R_{c,k}^{(i)})}{\log_2 M} +1 \right) + \frac{\mu_1 R_{c,k}}{\log_2 M} + \mu_2 \label{convex_u5}
 \end{equation}
\end{figure*}
\vspace{-0.4cm}
\subsection{Practical Coding Case}
Similarly, we also utilize the SCA method to approximately solve Problem (P4) for the practical coding case. We introduce auxiliary variables $ p_k = \ln P_k $,  $ r_k = \log_{10} R_{c,k} $, $ \hat{\rho}_{b,k} = \log_{10} \rho_{b,k} $, $ d_{o,k} $ and $ d_{s,k} $,  $ \forall  k \in\mathcal{K}$, and denote $ \bm{\hat{\nu}}_k = [p_k,r_k, \hat{\rho}_{b,k}, d_{o,k}, d_{s,k},  P_k, R_{c,k}] $ as the new design variable vector. Then,  
 Problem (P4) for the practical coding case is transformed as  
      \setcounter{equation}{50}
\begin{align}
   	\text{(P6.1)} \  \min\limits_{ \left\{ \bm{\hat{\nu}}_k\right\} } \  & \eqref{obj_simple_0}   \notag  \\
   \text{s.t.} \; \; & -\frac{1}{\sigma^2}|h_k|^2 e^{p_k + \frac{\lambda_1 R_{c,k}}{\log_2 M} + \lambda_2} +  \frac{\mu_1 R_{c,k}}{\log_2 M} + \mu_2 \notag \\  &\leq \hat{\rho}_{b,k},  \  \forall k \in \mathcal{K},   \label{rho_b_p_ineq}   \\ 
   & p_k \leq \ln P_k, \  \forall k \in \mathcal{K}, \label{pk_P}  \\
    & \eqref{log_r}-\eqref{new_d_si_ineq},\label{R_constraints_p}  \end{align} 
where \eqref{rho_b_p_ineq} and \eqref{pk_P} are obtained by introducing auxiliary variable $ p_k = \ln P_k $ into \eqref{final_ber_appro} and \eqref{obj_ori_simp}. Similar to Problem (P5.1), for the practical coding case, Problem (P6.1) is equivalent to  Problem (P4) except in the case where  $ R_{c,k} =0 $.  
%
    \begin{algorithm}[htbp]
\label{alg_2}
	\caption{SCA-Based Algorithm to Solve Problem (P5.2) \label{al_spca} }
	\label{abel_1} 
\begin{algorithmic}[1]
	\Require  $ L_{\text{max}}$, $ K $,  $ P_{\text{max}} $, $ R_{s} $, and $ \{h_1, \cdots, h_K \} $.
	\Ensure $ P_k^\star $ and $ R_{c,k}^\star $ for all  $ k \in \mathcal{K} $.
     \State Set the iteration number $ i = 1 $;
    \State Initialize the local point $ r_k^{(1)} $,  $\hat{\rho}_{b,k}^{(1)}$,  $d_{o,k}^{(1)}$, $d_{s,k}^{(1)}$, $P_k^{(1)}$, and $R_{c,k}^{(1)}$, $ \forall k \in\mathcal{K} $, for Problem (P5.2.$i$);
    \State \textbf{Repeat}
    \State \quad \ \  Solve Problem (P5.2.$i$) via optimization tools, e.g., CVX \cite{boyd2004}, and denote its optimal solution as $ r_k^{\star} $,  $\hat{\rho}_{b,k}^{\star}$,  $d_{o,k}^{\star}$, $d_{s,k}^{\star}$, $P_k^{\star}$, and $R_{c,k}^{\star}$, $ \forall k \in\mathcal{K} $ ;
    \State \quad \ \ Update the local point as $ r_k^{(i+1)} = r_k^{\star} $,  $\hat{\rho}_{b,k}^{(i+1)} = \hat{\rho}_{b,k}^{\star}$,  $d_{o,k}^{(i+1)}= d_{o,k}^{\star}$, $d_{s,k}^{(i+1)}= d_{s,k}^{\star}$, $P_k^{(i+1)}=P_k^{\star}$, and $R_{c,k}^{(i+1)}=R_{c,k}^{\star}$, $ \forall k \in\mathcal{K} $;
    \State \quad \ \ Set $ i = i+1 $; 
    \State \textbf{Until} the fractional decrease of the objective value of  Problem (P5.2) is below a threshold $ \epsilon >0 $. 

	\end{algorithmic}

\end{algorithm}
 
 Note that Problem (P6.1) is non-convex since \eqref{log_r},    \eqref{new_d_oi_ineq}, \eqref{new_d_si_ineq}, and  \eqref{rho_b_p_ineq} are all non-convex constraints. In the $ i $-th iteration ($i>0 $), consider the current value of the design variable vector in Problem (P5.2) is denoted as $\bm{\hat{\nu}}_k^{(i)} = [p_k^{(i)}, r_k^{(i)},d_{o,k}^{(i)},d_{s,k}^{(i)},\hat{\rho}_{b,k}^{(i)},P_k^{(i)},R_{c,k}^{(i)}] $, $ \forall k \in\mathcal{K} $. The convex upper bounds for the LHSs of \eqref{log_r},    \eqref{new_d_oi_ineq}, and \eqref{new_d_si_ineq} have already given in \eqref{convex_u2}, \eqref{convex_u3}, and \eqref{convex_u4}. Similarly, the global upper bound for the LHS of \eqref{rho_b_p_ineq} is obtained by replacing the concave function in \eqref{rho_b_p_ineq} with its first-order Taylor expansion, i.e.,
 \begin{equation}
 	-\frac{1}{\sigma^2}|h_k|^2 e^{p_k + \frac{\lambda_1 R_{c,k}}{\log_2 M} + \lambda_2} +  \frac{\mu_1 R_{c,k}}{\log_2 M} + \mu_2   \leq U^{(i)}_{5,k}(p_k, R_{c,k}),
 \end{equation} 
 where $ U^{(i)}_{5,k}(p_k, R_{c,k}) $ is given as \eqref{convex_u5}.

  By replacing the LHSs of \eqref{log_r}, \eqref{new_d_oi_ineq}, \eqref{new_d_si_ineq}, and \eqref{rho_b_p_ineq} with their convex upper bounds \eqref{convex_u2}, \eqref{convex_u3}, \eqref{convex_u4}, and \eqref{convex_u5}, respectively, we have the following convex subproblem in the $i$-th iteration, i.e.,
       \setcounter{equation}{55}
          \vspace{-0.2 cm} 
\begin{eqnarray}
   	\text{(P6.1.}i\text{)}  &\min\limits_{ \left\{ \bm{\hat{\nu}}_k\right\} } \  & \eqref{obj_simple_0}   \notag  \\  & \text{s.t.} & U^{(i)}_{5,k}( p_k, R_{c,k}) \leq \hat{\rho}_{b,k}, \ \forall k \in \mathcal{K}, \\
    &   & \eqref{R_bound}, \eqref{R_constraints}, \eqref{u2_ineq}-\eqref{u4_ineq}, \eqref{pk_P},
  \end{eqnarray}
  allowing it to be efficiently solved.  Finally, we iteratively solve Problem $\text{(P6.1.}i\text{)} $ to tackle Problem $\text{(P6.1)} $.


	  \vspace{-0.3 cm}  
\section{Experimental Results}

In this section, we provide some experimental results to validate the effectiveness of the proposed ASCC scheme over parallel Gaussian channels for image recovery and classification tasks.
The experiment setup is given as follows:
\begin{itemize}
	\item $\textbf{Datasets} $: To evaluate the performance of the proposed ASCC scheme, we conduct experiments from the CUB-$200$-$2011$ dataset, which contains $11,788$ images of birds across $200$ species. This dataset is divided into $5,994$ training images and $5,794$ testing images, with all images resized into $256 \times 256$ pixels.
	\item $\textbf{Model architecture} $:  We adopt the hyperprior-based DNN model\cite{balle2018variational} as the DNNs of the feature extraction function $ f_{\phi} $ and the feature recovery function $ g_{\theta} $, and employ Resnet-$152$ \cite{resnet} for the image classification task.  We build the look-up table composed of $ N =23 $ hyperprior-based DNN models, each trained on the CUB-$200$-$2011$ training samples, with Bpp values ranging from $0.02$ to $1.41$. 
	\item $\textbf{Channel settings} $: For the single Gaussian channel scenario, its channel gain is set as $ |h|^2 =1 $; For the parallel Gaussian channel scenario, we consider a case with $K=8$ channels, where each channel gain is sampled from an exponential distribution with mean $1$. The specific channel gains are generated by using a random seed of $\text{rng}(3)$ in MATLAB, resulting in the values $ (2.0748, 1.5739,$ $ 1.2348, 0.6717, 0.5964, 0.3451, 0.1132, 0.1095) $ listed in descending order. The noise power $ \sigma^2 $ for both single and parallel Gaussian channel scenarios is set to be $1$. 
	\item $\textbf{Channel codes} $: For random coding, we assume that the bit errors in the source bits follow an i.i.d. uniform distribution. When the optimal channel coding rate $ R_{c,k}^{\star} $ and transmitted power $ P_k^{\star} $ are derived for the $ k $-th parallel channel,  transmission errors are simulated by randomly flipping the source bits with BER computed by \eqref{appro_rho}.  For practical coding, we adopt a Polar code of length 256 with successive cancellation list (SCL) decoding and an LDPC code of length 4096 with belief propagation (BP) decoding (maximum 20 iterations). Both coding schemes are implemented by using the Sionna library \cite{hoydis2022sionna}.   In the single-channel scenario, the modulation type is QPSK for SNR values below $6$ dB and 16-QAM for higher values, applicable to both the Polar and LDPC codes.  In the parallel-channel scenario, the modulation type is set to 16-QAM for both two practical coding schemes.
	\item $\textbf{Baseline schemes} $: For comparison, we consider two baseline schemes for image transmission and recovery: the deep JSCC scheme \cite{8723589} and the conventional SSCC scheme. After image recovery, Resnet-$152$ is used for image classification. In the deep JSCC approach, images are directly mapped into analog symbols for transmission, with the training SNRs and ABRs matched with the testing SNRs and ABRs. For the SSCC scheme, images are first compressed by the BPG compression method, followed by encoding with practical channel coding schemes, e.g., a $c_1$-rate $(\lceil 256c_1\rceil, 256)$ Polar code and a $ c_2 $-rate $(\lceil 4096c_2\rceil, 4096)$ LDPC code. Moreover, in the parallel channel case, we apply the truncated channel inversion (TCI) power allocation strategy to the baseline schemes, which allocates the total power to the first $ K_0\in \mathcal{K} $ channels as\cite{cover1999elements} 
  \vspace{-0.2 cm}
 \begin{equation}
 	P_k = \begin{cases}
 		\frac{P_{\text{max}}}{|h_k|^2 \sum_{k=1}^{K_0} \frac{1}{ |h_k|^2}}, &k = 1,\cdots, K_0, \\
 		0,      &k = K_0+1, \cdots, K.        
 	\end{cases}
 \end{equation}
\end{itemize}

\subsection{Image Recovery Performance}
     \begin{figure}[t] 
\centering

\subfigure[ \label{fig_psnr_power_single}PSNR (dB) vs. SNR (dB) with ABR being $0.08$ ]{
\includegraphics[width=3in]{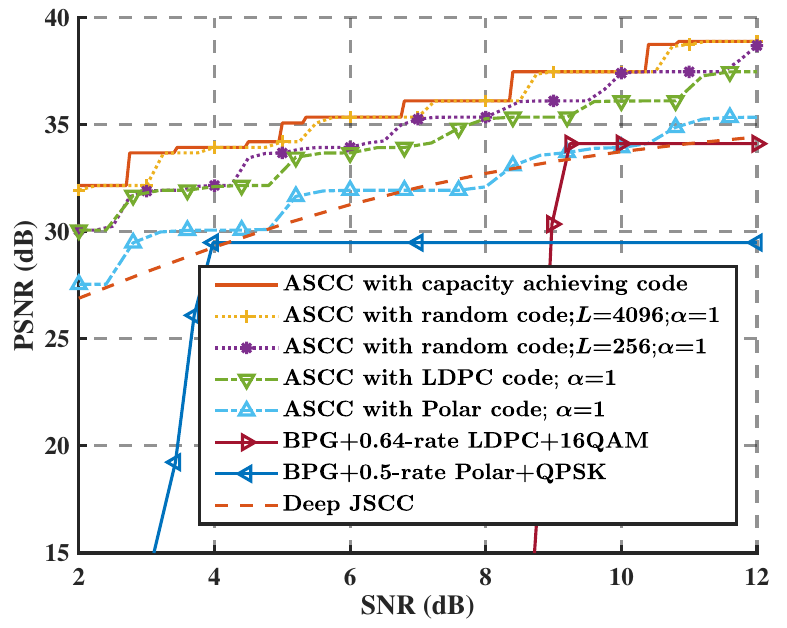}}
 \vspace{-0.3 cm}
 \subfigure[ \label{fig_single_bwr_PSNR}PSNR (dB) vs. ABR with SNR being $10$ dB ] {
\includegraphics[width=3in]{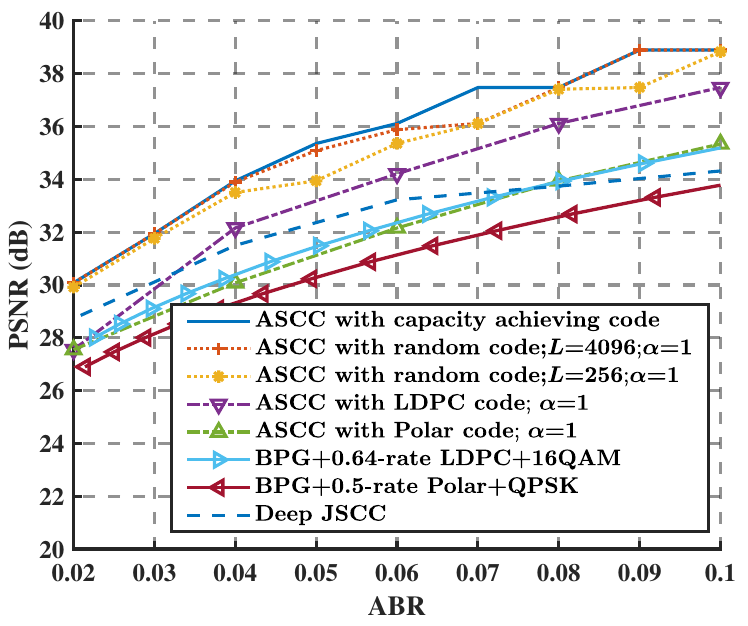}}

\caption{PSNR performance comparisons of the proposed ASCC scheme with baseline schemes over the single Gaussian channel.}
\label{fig_single_channnel_psnr}
\end{figure} 
 \vspace{-0.1 cm}

In this subsection, we utilize the peak signal-to-noise ratio (PSNR) metric\cite{balle2018variational} to quantify the image recovery performance of the proposed ASCC scheme, where the weight coefficient $ \alpha $ in \eqref{obj2} is set to be $1 $.   Fig. \ref{fig_psnr_power_single} depicts PSNR as a function of SNR over the single Gaussian channel with ABR being $0.08 $. It is observed that our proposed ASCC scheme exhibits a nearly stair-step increase in PSNR as SNR increases. This phenomenon can be explained as follows:  in the flatter regions, as the SNR increases, the channel decoder achieves near-perfect decoding, resulting in few channel errors. However, the increase in SNR is not yet sufficient to trigger the selection of a better DNN model from the look-up table, leading to limited gains in PSNR. In contrast, the steeply rising regions correspond to SNR levels where a better DNN model can be selected, significantly enhancing the reconstruction quality and resulting in a rapid increase in PSNR.  This phenomenon reveals the effectiveness of our proposed scheme in mitigating the cliff effect compared to the SSCC scheme. Furthermore, our proposed ASCC scheme outperforms nearly all baseline schemes, as it effectively adapts the source and channel coding rates to accommodate variations in SNR. For example, when the SNR is $ 10 $ dB,  the ``ASCC with LDPC code; $\alpha$$=$$1$'' scheme outperforms the ``Deep JSCC'' scheme and BPG schemes with LDPC code by around $ 2 $ dB and $ 2.1 $ dB, respectively, and the ``ASCC with random code; $\alpha$$=$$1$'' scheme outperforms the ``Deep JSCC'' scheme and BPG schemes with LDPC code by around $ 3 $ dB and $ 3.1 $ dB, respectively.   Notably, the ``ASCC with capacity achieving code'' scheme serves as a theoretical upper bound, representing the case where encoded bits are transmitted without errors at the capacity rate.

Next, Fig. \ref{fig_single_bwr_PSNR} shows the PSNR performance of the proposed ASCC scheme over the single Gaussian channel with respect to ABR, with the SNR being fixed at $ 10 $ dB.  It shows that the proposed ASCC scheme has superior performance in PSNR compared with all baseline schemes in most cases. For example, when the PSNR is $ 34 $ dB, the ``ASCC with LDPC code; $\alpha$$=$$1$'' scheme is able to save $ 0.02 \times M $ and $ 0.025 \times M $ bandwidths compared with the ``Deep JSCC'' scheme and the BPG scheme with LDPC coding, respectively.

\begin{figure*}  \label{fig_para_power_all}
\subfigure[PSNR (dB) vs. Total power (dBW) with ABR being $0.32$\label{fig_parallel_power_psnr}]{
\includegraphics[width=2.2in]{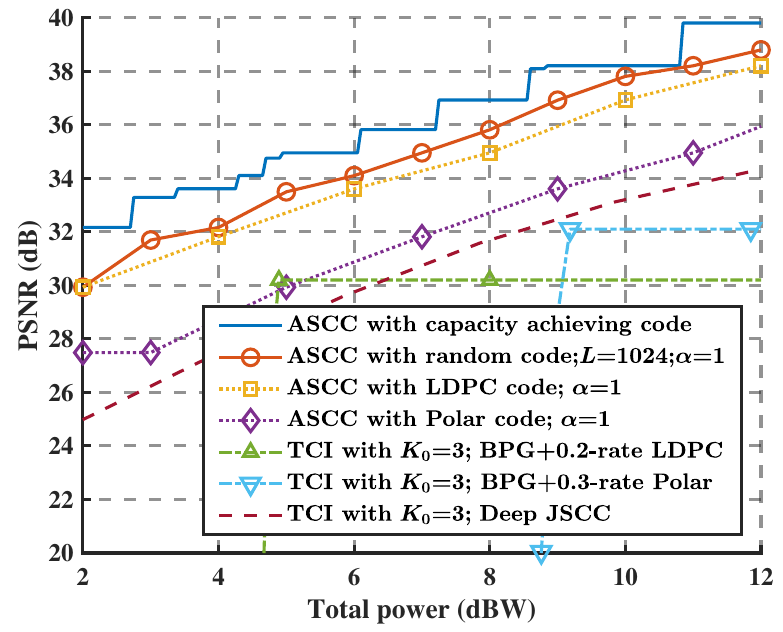}}
\hspace{0.1in}
\subfigure[Number of active channels vs. Total power (dBW) with ABR being $0.32$\label{fig_parallel_active_channels}]{
\includegraphics[width=2.1in]{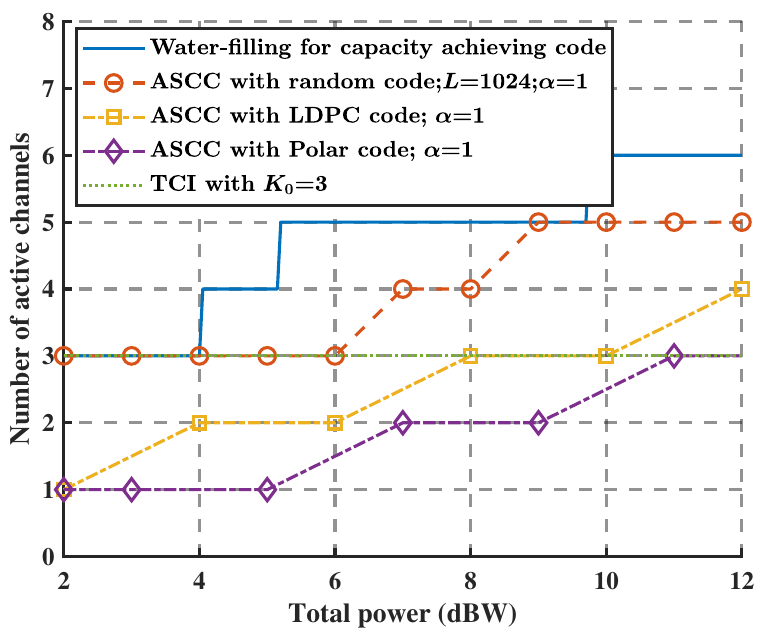}}
\hspace{0.1in}
\subfigure[PSNR (dB) vs. ABR with total power being $10$ dBW\label{fig_parallel_bwr_psnr}]{
\includegraphics[width=2.2in]{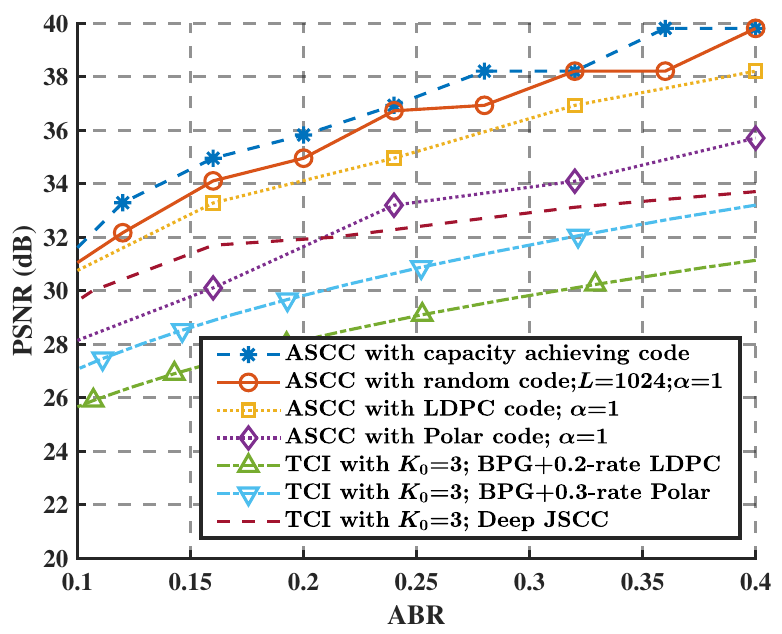}}

\caption{Image recovery performance of the proposed ASCC scheme and baseline schemes over the parallel Gaussian channels.}
\end{figure*}

Fig. 7 shows the image recovery performance over parallel Gaussian channels.  In Fig. \ref{fig_parallel_power_psnr}, the proposed ASCC scheme shows superior PSNR performance versus total power. Fig. \ref{fig_parallel_active_channels} illustrates how the number of active channels adapts with varying power levels. Fig.\ref{fig_parallel_bwr_psnr} shows PSNR versus ABR at fixed total power, confirming ASCC's advantage across different bandwidth ratios. The performance improvement can be attributed to the ability of the ASCC scheme to jointly optimize the source coding rate, power allocation, and channel coding rate, effectively minimizing the E2E distortion.
 
 \subsection{Classification Performance}
 
      \begin{figure}[htbp] 
\centering

\subfigure[ \label{fig_acc_power_single} Accuracy vs. SNR (dB) with ABR being $0.08$  ]{
\includegraphics[width=3in]{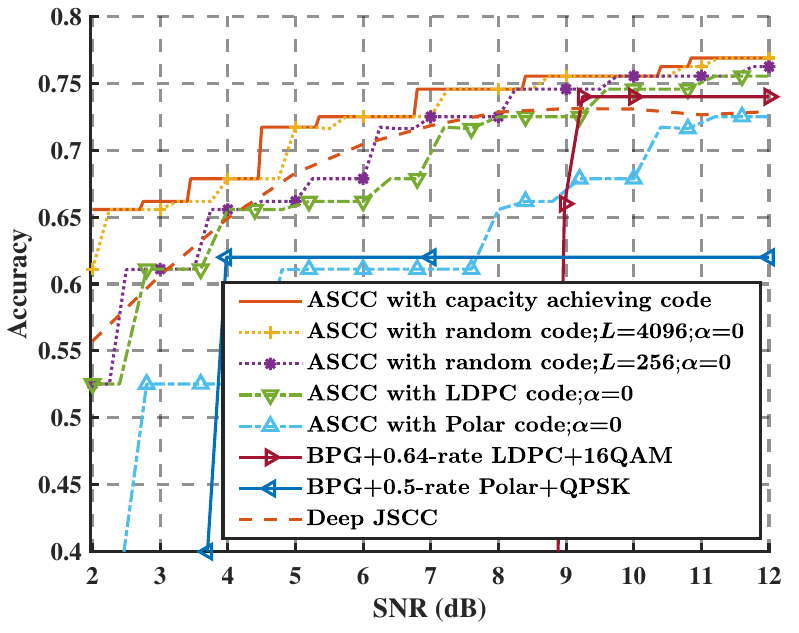}
} \hspace{0.51in}
\subfigure[ \label{fig_single_bwr_acc} Accuracy vs. ABR with SNR being $10$ dB]{
\includegraphics[width=3in]{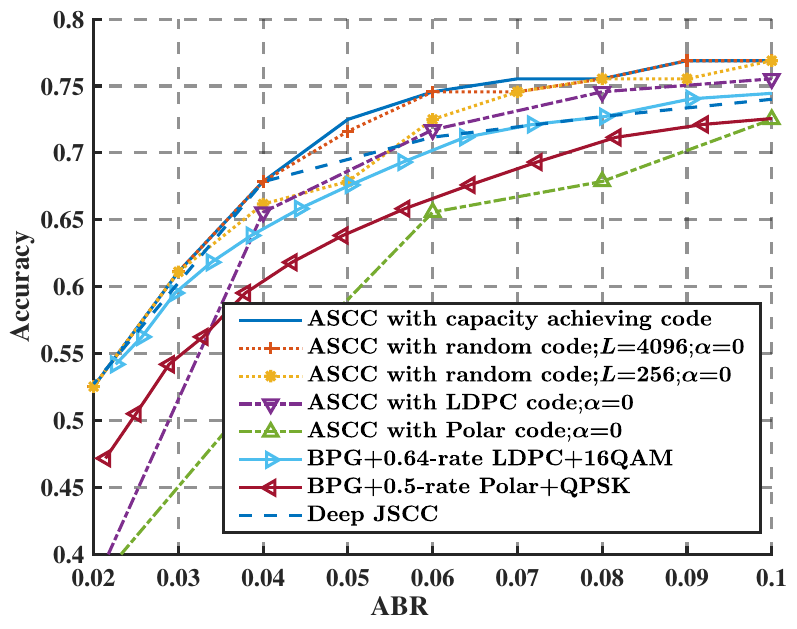}
}
\caption{Accuracy performance comparisons of the proposed ASCC scheme with the baseline schemes over the single Gaussian channel.}
\label{fig_single_channnel_acc}
\end{figure} 
 
 Then, we evaluate the classification performance of the proposed ASCC scheme using classification accuracy as the metric, where the weight coefficient $ \alpha $ in \eqref{obj2} is set to be $0 $.
 Fig. \ref{fig_acc_power_single} and Fig. \ref{fig_single_bwr_acc} plot the classification accuracy as functions of SNR and ABR over the single Gaussian channel, respectively. In  Fig. \ref{fig_acc_power_single}, it shows that the proposed ASCC scheme also exhibits a nearly stair-step increase in classification accuracy as SNR increases. Moreover, both figures show that the ASCC scheme has a comparable classification accuracy performance with the baseline schemes. For example, when SNR is around $ 8.2 $ dB, the classification accuracy of the ``ASCC with random code'' scheme is $ 2\% $ higher than that of the ``Deep JSCC'' scheme, while the classification accuracies of the ``ASCC with Polar code'' scheme and the ``BPG +$0.5$-rate Polar+QPSK'' scheme are quite lower than others since the codeword length of the Polar code is too small. Similarly,  as depicted in Fig. \ref{fig_single_bwr_acc}, when classification accuracy is $ 0.75 $, the ``ASCC with LDPC code'' scheme achieves this accuracy at an ABR of approximately 0.09, while all baseline schemes fail to reach the classification accuracy of $0.75$.

\begin{figure*}
\centering

\subfigure[Accuracy vs. Total power (dBW) with ABR being $0.32$\label{fig_parallel_power_acc}]{
\includegraphics[width=2.2in]{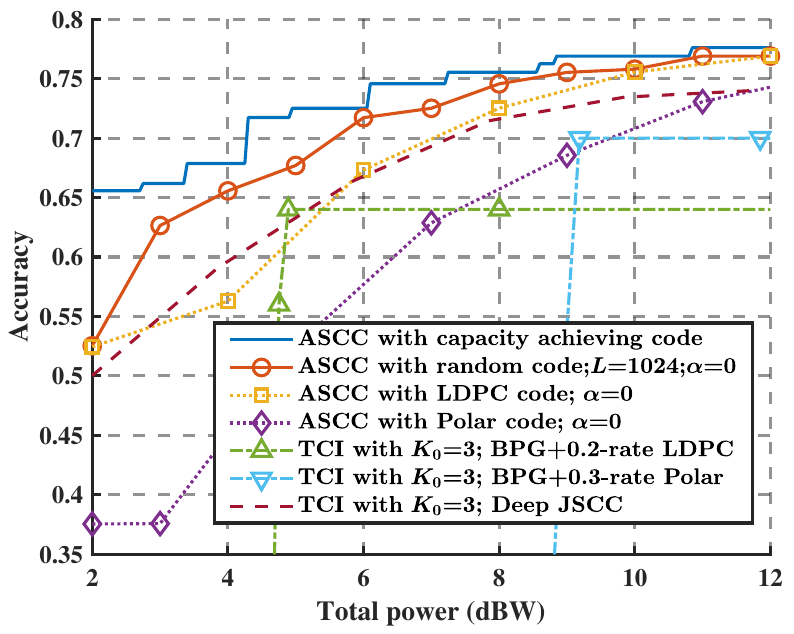}}
\hspace{0.1in}
\subfigure[Number of active channels vs. Total power (dBW) with ABR being $0.32$\label{fig_parallel_active_channels_0}]{
\includegraphics[width=2.1in]{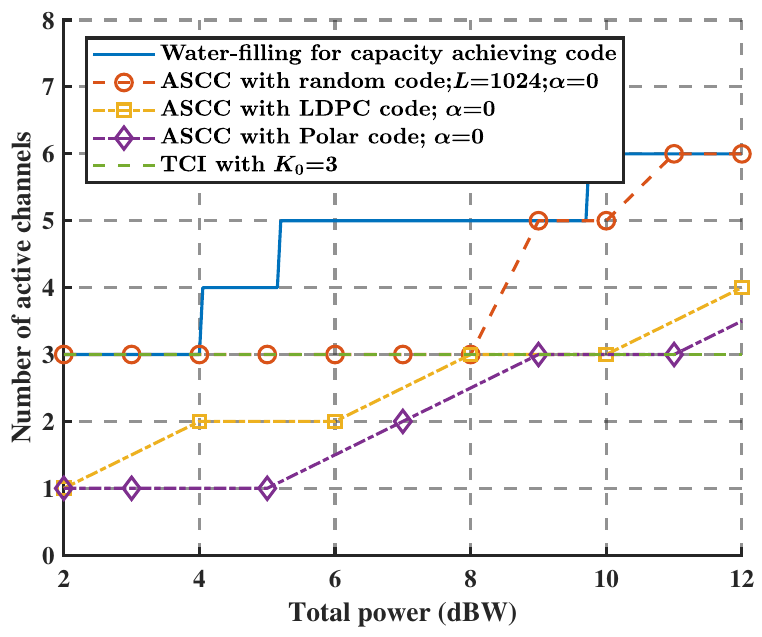}}
\hspace{0.1in}
\subfigure[Accuracy vs. ABR with total power being $10$ dBW\label{fig_parallel_bwr_acc}]{
\includegraphics[width=2.2in]{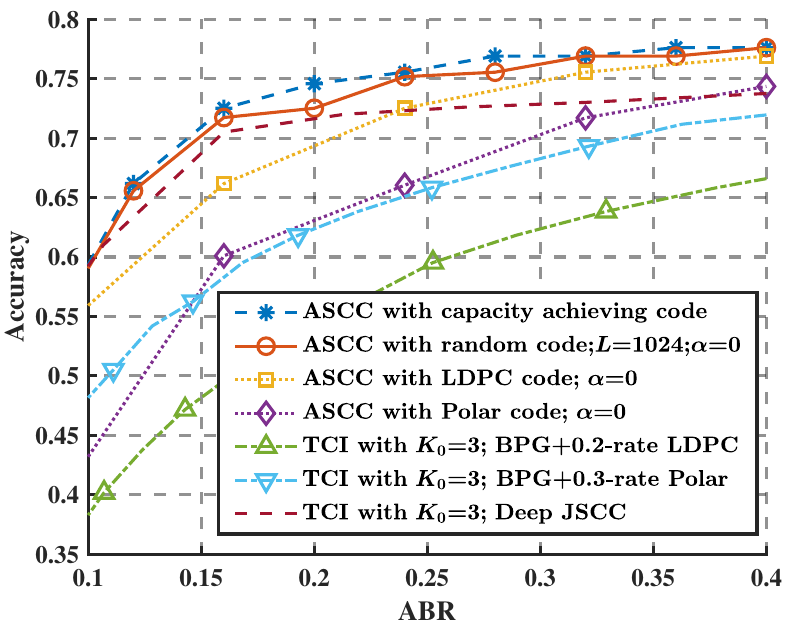}}

\caption{Classification accuracy performance of the proposed ASCC scheme with baselines over the parallel Gaussian channels.}
\label{fig_parallel_bwr}
\end{figure*}

Moreover, in the parallel-channel scenario, we depict the accuracy performance as a function of the total power budget in Fig. \ref{fig_parallel_power_acc}. It is observed that the ASCC scheme provides a significant improvement over the deep JSCC and BPG schemes. Fig. \ref{fig_parallel_active_channels_0} illustrates the corresponding channel allocation strategy, where the number of active channels for the ASCC scheme monotonically nondecreases with the total power budget. Then, Fig. \ref{fig_parallel_bwr_acc} shows classification accuracy as a function of ABR,   with the total power being $10$ dBW. The experimental results show that the proposed ASCC scheme with LDPC coding achieves higher PSNR compared with the deep JSCC and BPG schemes across the regions of ABR. Additionally, it outperforms both the deep JSCC and BPG schemes in classification accuracy when ABR exceeds $0.24$.

   \vspace{-0.3 cm}
\subsection{Joint Image Recovery and Classification}

      \begin{figure}[htbp] 
\centering

\subfigure[ \label{fig_psnr_alpha_0_1}PSNR (dB)   ]{
\includegraphics[width=1.68in]{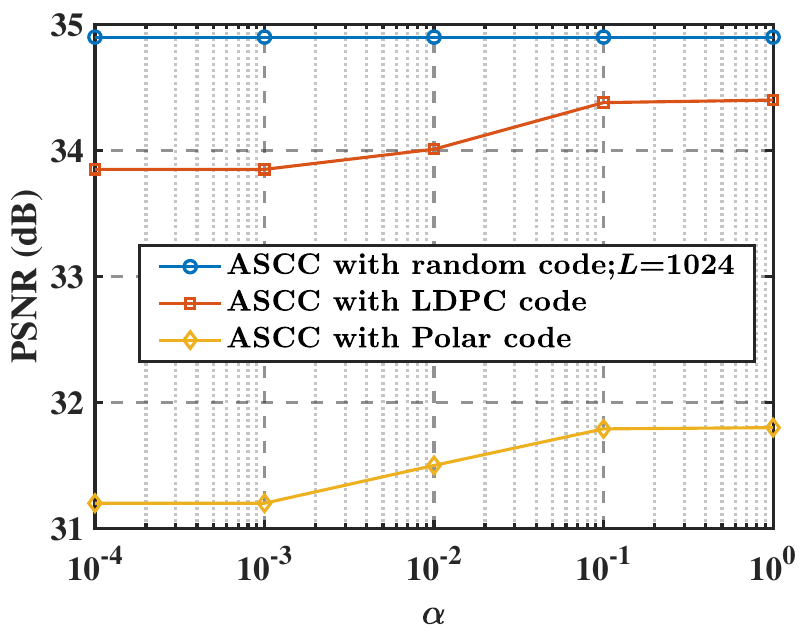}}
\hspace{-0.18in}
\subfigure[ \label{fig_acc_alpha_0_1} Accuracy]{
\includegraphics[width=1.71in]{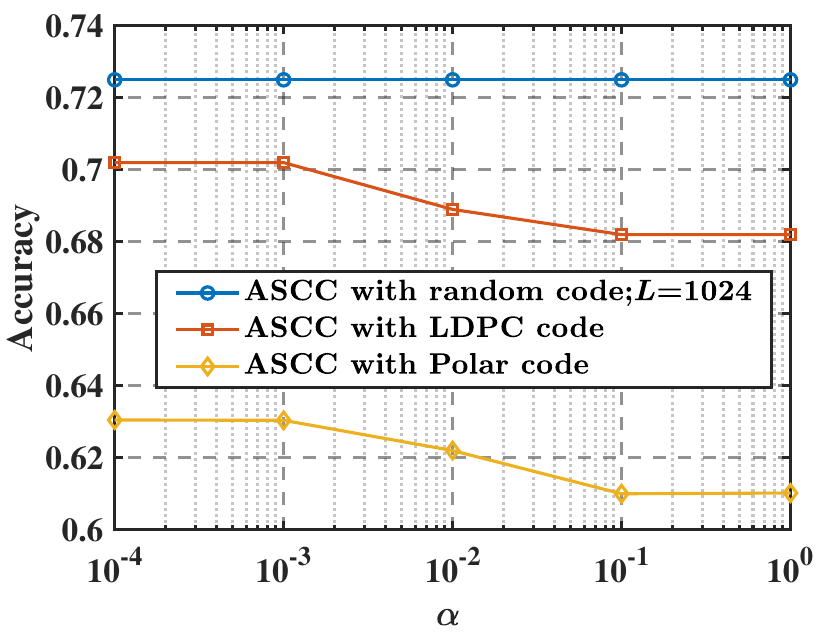}
}
\caption{PSNR and Accuracy performance versus weight coefficient $ \alpha $ over the parallel Gaussian channels, where the total power is $ 7 $ dBW, and the ABR is fixed at $0.32$.}
\label{fig_alpha_parallel}
\end{figure}  

Finally, we investigate the performance of the proposed ASCC scheme over parallel Gaussian channels as a function of the weight coefficient $ \alpha $ in $ D_\text{ave} $.  Fig. \ref{fig_psnr_alpha_0_1} shows that the PSNR performance improves slightly as $ \alpha $ increases, while Fig. \ref{fig_acc_alpha_0_1} indicates that the classification accuracy decreases as $ \alpha $ increases. This observation shows that $ \alpha $ serves as a trade-off parameter: smaller values prioritize classification accuracy, while larger values favor PSNR performance.
However, the choice of $ \alpha $ has a minor impact on the overall performance of the ASCC scheme. This phenomenon can be attributed to our cascaded network architecture at the receiver, where image recovery precedes classification. In this architecture, the classification and image recovery tasks are inherently coupled, where accurate classification largely depends on high-quality image reconstruction.
 A potential way to further analyze the impact of the weight coefficient $ \alpha $  on the ASCC scheme is to design two independent DNN-based decoders at the receiver: one for image reconstruction and the other for handling the classification task, which will be analyzed in future research.

  \vspace{-0.3 cm}
 \section{Concluding Remarks}
 
 This paper proposed an ASCC scheme for SemComs over parallel Gaussian channels that maintains compatibility with practical digital systems while enabling efficient adaptive transmissions. We first established explicit relationships between semantic source coding and digital channel coding by modeling E2E observation and semantic distortions as logistic functions of BER for any fixed source coding rate, where BER was further characterized as functions of SNR and channel coding rate for various channel coding schemes. Based on these distortion models, we developed a model selection-based optimization approach that jointly determines the optimal semantic source encoder from a pre-designed DNN look-up table and adapts channel coding rates and power allocation across parallel channels by using an SCA-based algorithm.
 Experimental results showed that the proposed ASCC exhibited superior performance gains compared with the deep JSCC and conventional SSCC schemes.

\appendices

 \vspace{-0.3 cm}
      \section{Proof of Proposition  \ref{Prpo_R}} \label{ap_prpo_R}

For the random coding case, the objective function in Problem (P3.1) is simplified as 
\begin{equation}
	O_r =  \frac{Q\left(\sqrt{\frac{L}{V(\gamma_{\text{max}})}} \left(C(\gamma_{\text{max}})-R_c\right)\ln2 \right) }{R_cL}.
\end{equation}
As proved in \cite{huang2024d}, the partial derivative of $ O_r $ with respect to channel coding rate $ R_c $ satisfies  $ \frac{\partial O_r }{\partial R_c } \geq 0  $ if $ R_c \geq \frac{\sqrt{2 \pi}\log_2 e}{2 \sqrt{L}} $ holds.  Therefore, the optimal solution to Problem (P3.1) is $ R^{\star}_c= \frac{R_{s}}{L_{\text{max}}} $ if $ \frac{R_{s}}{L_{\text{max}}} \geq \frac{\sqrt{2 \pi}\log_2 e}{2 \sqrt{L}}  $.

For the practical coding scheme, with fixed SNR and modulation order $ M $, we consider the case that $ \rho_b $ is monotonically nondecreasing as channel coding rate  $ R_c$ increases, which coincides with the observations in Remark \ref{Re_practical_code}. In this case, it is easy to see that the optimal solution to Problem (P3.1) is $ R^{\star}_c= \frac{R_{s}}{L_{\text{max}}} $. Therefore, we have completed this proof.
 \vspace{-0.3 cm}
 \section{Proof of Proposition \ref{Prop_total_rate} } \label{ap_Prop_total_rate}
 	We prove this proposition by contradiction. Suppose $ [R_1^\star, \cdots, R_{c,K}^\star,P_1^\star,\cdots, P_K^\star] $ is the optimal solution to Problem (P2) satsifying $ \sum_{k=1}^K R_{c,k}^\star >  \frac{KR_{s}}{L_{\text{max}}}$. For this solution, $ \rho_{b,k}^\star $ is obtained by \eqref{appro_rho}.  
 	Define $ \eta \triangleq \frac{KR_{s}/L_{\text{max}}}{\sum_{k=1}^K R_{c,k}^\star} $ and construct a new solution with $ R_{c,k}^\prime \triangleq \eta R_{c,k}^\star $, $ \forall k \in \mathcal{K} $. Note that $ \eta < 1 $ under our assumption. Let $ \rho_{b,k}^\prime $ denote the value of $ \rho_{b,k}$ when $ R_{c,k} = R_{c,k}^{\prime} $ and $ P_k = P_k^{\star} $ in \eqref{appro_rho}. Since $ \rho_{b,k} $ is monotonically increasing with respect to $ R_{c,k} $ shown in Appendix \ref{ap_prpo_R}, we have $ \rho_{b,k}^\prime < \rho_{b,k}^\star $. 
 Then, the objective function in Problem (P2) satisfies 
 	\begin{align}
 		&\sum_{k=1}^{K}\frac{R_{c,k}^\star\left( \alpha 10^{\hat{D}_o(R_s,\rho_{b,k}^\star)} + (1-\alpha)D_s(R_s,\rho_{b,k}^\star) \right)}{\sum_{k^{\prime}=1}^{K}R_{c,k^{\prime}}^\star} \notag \\ = &\sum_{k=1}^{K}\frac{R_{c,k}^\prime\left( \alpha 10^{\hat{D}_o(R_s,\rho_{b,k}^\star)} + (1-\alpha) D_s(R_s,\rho_{b,k}^\star) \right)}{\sum_{k^{\prime}=1}^{K}R_{c,k^{\prime}}^\prime}, \label{ap_R_prime} \\ >&\sum_{k=1}^{K}\frac{R_{c,k}^\prime\left( \alpha 10^{\hat{D}_o(R_s,\rho_{b,k}^\prime)} + (1-\alpha) D_s(R_s,\rho_{b,k}^\prime) \right)}{\sum_{k^{\prime}=1}^{K}R_{c,k^{\prime}}^\prime}, \label{ap_R_prime2}
 	\end{align}
 where \eqref{ap_R_prime} follows from the definition of $ R_{c,k}^\prime $ and \eqref{ap_R_prime2} holds since  $ \hat{D}_o $ and $ D_s $ are both monotonically increasing as $ \rho_{b,k} $ increases. Moreover, from \eqref{ap_R_prime2}, it reveals that $ [R_1^\star, \cdots, R_{c,K}^\star,P_1^\star,\cdots, P_K^\star] $ is no longer the optimal solution to Problem (P2), which makes a contradiction. Therefore, we have proved this proposition. 
  
  \vspace{-0.3 cm}
    \section{Proof of Proposition \ref{Prop_up}} \label{ap_Prop_up} 
  
  To prove this proposition, we first show that $ Q $-function satisfies the following two inequalities.
  \begin{Lemma} \label{ap_le_Q}
  $ e^{-\frac{w^2}{2}} - \sqrt{2\pi}wQ(w)\geq 0 $	holds for all $ w \in \mathbb{R} $.
  \end{Lemma}
  \begin{IEEEproof}
   To facilitate this proof, we first define a new function as $ g(w)\triangleq e^{-\frac{w^2}{2}} - \sqrt{2\pi}wQ(w) $.  The first derivative of $ g(w) $ satisfies $ g\prime(w) = - \sqrt{2\pi}Q(w) < 0 $, which reveals that $ g(w) $ is  monotonically decreasing with respect to $ w$. Moreover, it is easy to obtain $ \lim_{w \to \infty} g(w) =0$ by L'H\^opital's Rule. Therefore, we have proved this lemma.
  \end{IEEEproof}
    \begin{Lemma} \label{ap_Le_Q_ineq}
	For all $ w, \hat{w} \in \mathbb{R} $, $ Q(w) $ is upper bounded by  
	\begin{equation}
		Q(w) \leq Q(\hat{w})e^{-\hat{a}(\hat{w})(w-\hat{w})}, \label{ap_Q_upper}
	\end{equation} 
	with $ \hat{a}(\hat{w}) $ being $  \hat{a}(\hat{w}) = e^{-\frac{\hat{w}^2}{2}}/(\sqrt{2\pi}Q(\hat{w}))$.  Besides, the equality in \eqref{ap_Q_upper} holds when $ w= \hat{w} $.
\end{Lemma}
\begin{IEEEproof}
	As proved in \cite{9685691}, for any real number $\hat{w} $, $ Q $-function satisfies 
  \begin{equation}
  	Q(w) \leq \hat{b}(\hat{w})e^{-\hat{a}(\hat{w})w}+\hat{c}(\hat{w}), \label{ap_Q_ineq}
  \end{equation}
   for all $ w \in \mathbb{R} $ and the equality holds when $ w= \hat{w} $, where $ \hat{a}(\hat{w})  $, $ \hat{b}(\hat{w}) $, and $ \hat{c}(\hat{w}) $ are defined as $ \hat{a}(\hat{w}) \triangleq  \max\{ \frac{e^{-\frac{\hat{w}^2}{2}}}{\sqrt{2\pi}Q(\hat{w})}, \hat{w} \} $, $ \hat{b}(\hat{w}) \triangleq  \frac{1}{\sqrt{2\pi}\hat{a}(\hat{w})}e^{\hat{a}(\hat{w})\hat{w}-\frac{\hat{w}^2}{2}} $, and $ \hat{c}(\hat{w})\triangleq  Q(\hat{w})- \hat{b}(\hat{w})e^{-\hat{a}(\hat{w})\hat{w}}$, respectively. From Lemma \ref{ap_le_Q}, it is easy to see that $\hat{a}(\hat{w}) $ can be simplified as $  \hat{a}(\hat{w}) = e^{-\frac{\hat{w}^2}{2}}/(\sqrt{2\pi}Q(\hat{w}))  $. Accordingly, $ \hat{b}(\hat{w})$ and $ \hat{c}(\hat{w}) $ are simplifed as $ \hat{b}(\hat{w}) = Q(\hat{w}) e^{\hat{a}(\hat{w})\hat{w}} $ and $  \hat{c}(\hat{w}) = 0 $, respectively. Therefore, \eqref{ap_Q_ineq} is simplified as \eqref{ap_Q_upper}, and we complete this proof. 
\end{IEEEproof}

Finally, the upper bound $ U^{(i)}_{1,k}( P_k, R_{c,k}) $ in \eqref{upper1} is obtained by setting $ w = \sqrt{L} \left(C(\gamma_k)-R_{c,k}\right)\ln2 $ and $ \hat{w} =\hat{w}_{k}^{(i)}  $ in \eqref{ap_Q_upper}, which can be easily checked to be a convex function.

   \bibliographystyle{IEEEtran}   
  \bibliography{semantic_point_to_point_twc_v1_clear_arxiv} 
  
\end{document}